\documentstyle[11pt,aasms4]{article}

\newcommand{\mgfld}{{\small MGFLD}}
\newcommand{\ppm}{{\small PPM}}
\newcommand{\nes}{{\small NES}}

\newcommand{\be}{\begin{eqnarray}}
\newcommand{\ee}{\end{eqnarray}}

\newcommand\ie {{\it i.e.}}
\newcommand\eg {{\it e.g.}}
\newcommand\etal{{\it et al.}}

\newcommand\viz{{\it viz.}}
\newcommand\gcm{g~cm$^{-3}$}
\newcommand\simgreater{\,\lower0.7ex\hbox{$\stackrel{>}{\sim}$}\,}
\newcommand\simless{\,\lower0.7ex\hbox{$\stackrel{<}{\sim}$}\,}
\newcommand\msol{${\rm M}_\odot$}

\newcommand{\nue}{$\nu_{{\rm e}}$}
\newcommand{\nuebar}{$\bar{\nu}_{{\rm e}}$}

\begin{document}

\title{The Interplay Between Protoneutron Star Convection and\\ Neutrino Transport
in Core Collapse Supernovae}

\author{A. Mezzacappa\altaffilmark{1,2}, 
A. C. Calder\altaffilmark{1,3}, 
S. W. Bruenn\altaffilmark{4}, 
J. M. Blondin\altaffilmark{5},\\ 
M. W. Guidry\altaffilmark{1,2},
M. R. Strayer\altaffilmark{1,2}, and 
A. S. Umar\altaffilmark{3}}
\vspace{4.0in}
\begin{center}
Accepted for publication in {\it The Astrophysical Journal}
\end{center}

\altaffiltext{1}{Theoretical and Computational Physics Group, Oak Ridge National
Laboratory, Oak Ridge, TN 37831--6354}
\altaffiltext{2}{Department of Physics and Astronomy, University of Tennessee,
Knoxville, TN 37996--1200}
\altaffiltext{3}{Department of Physics and Astronomy, Vanderbilt University,
Nashville, TN 37235}
\altaffiltext{4}{Department of Physics, Florida Atlantic University, Boca Raton, FL 33431--0991}
\altaffiltext{5}{Department of Physics, North Carolina State University, 
Raleigh, NC 27695--8202}
\newpage

\begin{center}
{\large {\bf Abstract}}
\end{center}

We couple two-dimensional hydrodynamics to realistic one-dimensional
multigroup flux-limited diffusion neutrino transport to investigate
protoneutron star convection in core collapse supernovae, and more 
specifically, the interplay between its development and neutrino 
transport. Our initial conditions, time-dependent boundary conditions, 
and neutrino distributions for computing neutrino heating, cooling, 
and deleptonization rates are obtained from one-dimensional simulations 
that implement multigroup flux-limited diffusion and one-dimensional 
hydrodynamics.

The development and evolution of protoneutron star convection are 
investigated for both 15 and 25 \msol\ models, representative of  the
two classes of stars with compact and extended iron cores, 
respectively. For both models, in the absence of neutrino transport, 
the angle-averaged radial and angular convection velocities in the 
initial Ledoux unstable region below the shock after bounce achieve 
their peak values in $\sim$ 20 ms, after which they decrease as the 
convection in this region dissipates. The dissipation occurs as the 
gradients are smoothed out by convection. This initial protoneutron 
star convection episode seeds additional convectively unstable
regions farther out beneath the shock. The additional 
protoneutron star convection is driven by successive
negative entropy gradients that develop as the shock, in propagating
out after core bounce, is successively strengthened and weakened by
the oscillating inner core. The convection beneath the shock distorts
its sphericity, but on the average the shock radius is not boosted
significantly relative to its radius in our corresponding
one-dimensional models. 

In the presence of neutrino transport, protoneutron star convection 
velocities are too small relative to bulk inflow velocities to result 
in any significant convective transport of entropy and leptons. This 
is evident in our two-dimensional entropy snapshots, which in this 
case appear spherically symmetric. The peak angle-averaged radial 
and angular convection velocities are orders of magnitude smaller 
than they are in the corresponding ``hydrodynamics only'' models. 

A simple analytical model supports our numerical results, indicating 
that the inclusion of neutrino transport reduces the entropy-driven 
(lepton-driven) convection growth rates and asymptotic velocities by 
a factor $\sim$ 3 (50) at the neutrinosphere and a factor $\sim$ 250 
(1000) at $\rho=10^{12}$ g/cm$^{3}$, for both our 15 and 25 \msol\ models. 
Moreover, when transport is included, the initial postbounce entropy 
gradient is smoothed out by neutrino diffusion, whereas the initial 
lepton gradient is maintained by electron capture and neutrino escape 
near the neutrinosphere. Despite the maintenance of the lepton gradient, 
protoneutron star convection does not develop over the 100 ms duration 
typical of all our simulations, except in the instance where ``low-test'' 
initial conditions are used, which are generated by core collapse and 
bounce simulations that neglect neutrino--electron scattering and 
ion--ion screening corrections to neutrino--nucleus elastic scattering.

Models favoring the development of protoneutron star convection either 
by starting with more favorable, albeit artificial (``low-test''), initial 
conditions or by including transport corrections that were ignored in our 
``fiducial'' models were considered. Our conclusions nonetheless remained 
the same. Evidence of protoneutron star convection in our two-dimensional 
entropy snapshots was minimal, and as in our ``fiducial'' models, the 
angle-averaged convective velocities when neutrino transport was included 
remained orders of magnitude smaller than their counterparts in the corresponding 
``hydrodynamics only'' models. 

\newpage

\section{Introduction}
\bigskip

\subsection{Current Core Collapse Supernova Paradigm}

The current core collapse supernova paradigm begins with the collapse of the iron core of an
evolved massive star at the end of its thermonuclear evolution. The rebound of the inner core,
brought about by the rapid increase of pressure with density when the density rises above
nuclear matter density, generates a shock wave at a radius $\sim 20$ km, and drives it into 
the outer core. All groups currently agree that, rather than propagating directly outward 
through the entire envelope and producing a supernova, the shock stalls and turns into an 
accretion shock at a radius of 100 -- 200 km. This happens for several reasons: (1) Nuclear
dissociation behind the shock lowers the ratio of pressure to energy, thereby lowering the
strength of the shock for a given energy delivered to it by the inner core. (2) There is a
reduction in both the thermal and lepton number contribution to the shock pressure as a
result of the intense burst of \nue\ radiation through the shock that ensues when it 
passes outwardly through the \nue-sphere, which at this time is located at $\sim 60$ km.

All groups also currently believe that the supernova is powered by some variant of the ``shock
reheating mechanism'' of Wilson (1985) and Bethe and Wilson (1985). In this mechanism, the
material behind the shock heats on a time scale of hundreds of milliseconds by the transfer 
and deposition of energy by neutrinos from the hot contracting core. As a result of this
energy transfer, the shock is once again driven out. This is in contrast to the original 
neutrino-driven supernova mechanism of Colgate and White (1966), which operated on the 
dynamical time scale of milliseconds. 

The energy transfer between neutrinos and matter behind the shock is mediated primarily by 
the charged current reactions $\nu_{\rm e} + {\rm n} \rightleftharpoons {\rm p} + {\rm e^{-}}$ 
and $\bar{\nu}_{\rm e} + {\rm p} \rightleftharpoons {\rm n} + {\rm e^{+}}$. When these
reactions proceed to the right, the matter heats; when they proceed to the left, the matter
cools. To drive these reactions sufficiently rapidly to the right for a supernova to develop
requires a critical combination of \nue-\nuebar\ luminosity and {\small RMS} energy (\eg, Burrows and
Goshy 1993) that no recent realistic, spherically symmetric, numerical simulation has to date 
delivered (Bruenn 1993; Cooperstein 1993; Wilson and Mayle 1993).

\subsection{Role of Fluid Instabilities}

A possible way out of this impasse is illuminated by the early recognition that neutrino 
luminosities and {\small RMS} energies might be enhanced by fluid instabilities occurring below 
the neutrinosphere (Epstein 1979; Bruenn, Buchler, and Livio 1979; Colgate and Petschek 
1980; Mayle 1985; Wilson \etal\ 1986; Arnett 1987; Burrows 1987; Burrows and Lattimer 
1988), or that the neutrino energy deposition efficiency might be increased by fluid 
instabilities occurring above the neutrinosphere (Bethe 1990; Herant, Benz, and Colgate 
1992; Colgate, Herant, and Benz 1993). Both of these possibilities are being vigorously 
pursued by a number of groups. Additional motivation for the study of fluid instabilities 
at this early stage in the development of a supernova is the existence of a large variety 
of observational hints that large-scale mixing processes occurred in the early phase of 
the explosion of SN 1987A, and played an important role in generating the observables 
(\eg, see Herant, Benz, and Colgate 1992, 1994; Burrows, Hayes, and Fryxell 1995; Janka and
M\"{u}ller  1996).

\subsection{Fluid Instabilities below the Neutrinosphere}

Consider the region in the vicinity of and below the neutrinosphere, which will be the
subject of this paper. Material in this region is in nuclear statistical equilibrium, and
its thermodynamic state and composition can therefore be specified by just three independent
variables. We will choose these to be the dimensionless entropy per baryon, $s$, the lepton
fraction, $Y_{\ell}$, and the pressure, $P$. Above the neutrinosphere, electrons and neutrinos
are no longer strongly interacting and $Y_{\ell}$ should be replaced by the electron fraction,
$Y_{\rm e}$. In the presence of gradients in $s$ and/or $Y_{\rm \ell}$, a strong gravitational
field, and the neutrino transport of energy and leptons, the material may be unstable to any one
of a number of fluid instabilities (Bruenn and Dineva 1996). If we ignore the effects of
neutrino transport for the moment, the situation simplifies and the number of possible fluid
instabilities is reduced to one, \viz, the Rayleigh-Taylor (convective) instability. Under
these conditions, the criterion for convective instability is the ``Ledoux'' condition, given by

\begin{equation}
\left( \frac{ \partial \rho }{ \partial \ln Y_{\ell} } \right)_{s,P}
\left( \frac{ \partial \ln Y_{\ell} }{ \partial r } \right)
+  \left( \frac{ \partial \rho }{ \partial \ln s } \right)_{Y_{\ell},P}
\left( \frac{ \partial \ln s }{ \partial r } \right) > 0
\end{equation}
\smallskip

\noindent The derivative $\left( \frac{ \partial \rho }{ \partial \ln s } \right)_{Y_{\ell},P}$ is
negative for all thermodynamic states; therefore, negative entropy gradients always tend to be
destabilizing. On the other hand, the derivative $\left( \frac{ \partial \rho }{ \partial \ln Y_{\ell} }
\right)_{s,P}$ is negative for large $Y_{\ell}$, but becomes positive for small 
$Y_{\ell}$.\footnote{An increase in $Y_{\ell}$ implies an increase in the number of particles, 
and this must be accompanied by a decrease in the temperature $T$ to keep $s$ constant. 
The increase in particle number tends to increase $P$, while the decrease in temperature tends
to decrease $P$. At large $Y_{\ell}$, the increase in the number of particles has the more 
important effect on $P$, and $\rho$ must decrease to keep $P$ constant. Under these 
conditions, the derivative $\left( \frac{ \partial \rho }{ \partial \ln Y_{\ell} }
\right)_{s,P}$ is therefore negative. At small $Y_{\ell}$, the decrease in 
temperature is the more important effect, and $\rho$ must increase to keep $P$ constant, making  
the derivative positive.}  The 
critical value of $Y_{\ell}$ at which the derivative changes sign depends on $s$ and $P$, but 
is typically in the range 0.1 -- 0.2. Thus, negative gradients in $Y_{\ell}$ initially tend to 
be destabilizing, but become stabilizing when the core deleptonizes sufficiently for values of 
$Y_{\ell}$ to go below the critical value. We will refer to convection driven by negative gradients 
in $s$ and $Y_{\ell}$ near and below the neutrinosphere as ``entropy-driven protoneutron star 
convection'' and ``lepton-driven protoneutron star convection,'' respectively. These modes of 
convection are to be distinguished from the entropy-driven convection that occurs above the 
neutrinosphere after shock stagnation. This convection is sustained by neutrino heating, and 
will be referred to as ``neutrino-driven convection.'' It will be the subject of another paper 
(Mezzacappa \etal\ 1997; see also, Herant \etal\ 1992, 1994, Miller et al. 1993, Burrows \etal\ 
1995, and Janka and M\"{u}ller 1996).

If we now bring the neutrino transport of energy and leptons into the picture, entropy- and
lepton-driven protoneutron star convection are still possible (with reduced growth rates), and
two additional modes of instability, ``neutron fingers'' and ``semiconvection,'' become possible.
[See Bruenn and Dineva (1996) for a physical description of these instabilities.] These instabilities 
occur on a diffusion rather than on a dynamical time scale, and usually require that one of the 
gradients ($s$ or $Y_{\ell}$) be stabilizing while the other be destabilizing, and that energy 
and lepton equilibration by neutrinos proceed at different rates.

There are several reasons why fluid instabilities in the region below the neutrinosphere
may play an important role in the shock reheating mechanism. All have to do with the fact that
instabilities tend to drive fluid motions that tend to circulate the fluid throughout the
unstable region. The fluid motions associated with entropy-driven protoneutron star convection
will tend to advect high-entropy material from deeper in the core to the vicinity of the
neutrinosphere, increasing the temperature of the neutrinosphere and thereby increasing
both the \nue\ and the \nuebar\ emission rates. Lepton-driven protoneutron star convection 
will tend to advect lepton-rich material from deeper in the core to the vicinity of the
neutrinosphere, increasing and decreasing the \nue\ and \nuebar\ emission rates, respectively,
and hastening the deleptonization of the core.

\subsection{Lepton-Driven Convection}

As we will indicate below, if protoneutron star convection is important for shock reheating, 
lepton-driven protoneutron star convection is probably more important than entropy-driven
protoneutron star convection. A negative lepton gradient results from the fact that
the region well below the neutrinosphere is lepton rich relative to the material flowing inward
through the neutrinosphere at the time of shock stagnation. This is because the latter
is completely shock dissociated into free neutrons and protons and is hydrostatically
settling onto the core when it passes through the neutrinosphere. Electron capture therefore
proceeds rapidly on the plethora of free protons, and plenty of time is available for the \nue's
produced by these electron captures to escape. Thus, the material in the vicinity of the
neutrinosphere at this time is very lepton poor ($Y_{\ell} \simless 0.1)$. On the other
hand, material deep within the core passed rapidly through the neutrinosphere  during infall, and
either never encountered the shock, or encountered the shock below the neutrinosphere, as the shock
was beginning to form. This material was therefore ``cold,'' and contained only a small mass
fraction of free protons when it passed inwardly through the neutrinosphere. As a result,
electron capture proceeded more slowly and the \nue's produced by the electron capture had
less time to escape. Consequently, the material residing deep within the core has a higher
lepton fraction ($Y_{\ell} \sim 0.3)$ at the time of shock stagnation. A negative gradient in
$Y_{\ell}$ below the neutrinosphere is thus naturally established after shock stagnation, and
this gradient is maintained by continuing deleptonization near the neutrinosphere.

The negative gradient in $Y_{\ell}$, unless stabilized by a positive gradient in $s$, will
drive a convective flow that may advect leptons from the lepton-rich interior to 
the vicinity of the neutrinosphere, where they will be radiated away as \nue's. This is the 
basis of the original suggestion of Epstein (1979), and the early work of Bruenn, Buchler, 
and Livio (1979), Colgate and Petschek (1980), and Livio, Buchler and Colgate (1980). Interest
in this early work waned when it was pointed out that the positive entropy gradient established
by reenergizing the shock would stabilize much of this region (Lattimer and Mazurek 1981;
Smarr \etal\ 1981), and because of the growing interest in the ``prompt'' supernova mechanism. 
However, as the prompt mechanism proved untenable and attention turned to shock reheating, and 
as SN 1987A appeared with its many hints of fluid instabilities occurring early on, interest
in the role of fluid instabilities in the supernova mechanism revived 
(
Arnett 1986, 1987; 
Bethe 1990, 1993; 
Bruenn and Mezzacappa 1994; 
Bruenn, Mezzacappa, and Dineva 1995; 
Bruenn and Dineva 1996; 
Burrows 1987; 
Burrows and Lattimer 1988; 
Burrows and Fryxell 1992, 1993; 
Burrows, Hayes and Fryxell 1995; 
Colgate, Herant, and Benz 1993; 
Herant, Benz, and Colgate 1992; 
Herant \etal\ 1994; 
Janka and M\"{u}ller 1993a, 1993b, 1995, 1996; 
M\"{u}ller 1993; 
M\"{u}ller and Janka 1994;
Shimizu, Yamada, and Sato 1993, 1994; 
Wilson and Mayle 1988, 1993 
).

\subsection{Neutron Fingers}

The stabilizing effect of the positive entropy gradient established by the shock as it gathers
power from the rebounding inner core prompted Mayle and Wilson to invoke neutron fingers as
a driver of fluid circulation below the neutrinosphere (Mayle 1985, Wilson \etal\ 1986, Wilson
and Mayle 1988, 1993). Modeling the fluid circulation by a mixing-length algorithm, they
found that if the fluid circulation were made sufficiently vigorous, they could generate
explosions out of what would otherwise have been duds. Neutron fingers can occur in the
presence of a positive entropy gradient and a negative lepton gradient if the energy
equilibration rate of a fluid element with the background is faster than the lepton
equilibration rate. The opposite was found to be the case, however, in recent
investigations by Bruenn, Mezzacappa and Dineva (1995), and Bruenn and Dineva (1996). The
implication of these investigations is that the material, if unstable at all, will be unstable to 
semiconvection rather than neutron fingers. Semiconvection tends to drive more localized 
flows and consequently may not transport leptons efficiently to the neutrinosphere.

\subsection{Entropy-Driven Convection}

While the shock establishes a positive entropy gradient as it gathers strength, Arnett (1986, 1987)
pointed out that the dissipation of the shock farther out, due to nuclear dissociation and \nue\
radiation, will imprint a negative entropy gradient and therefore destabilize this region to
entropy-driven convection. While Arnett (1986, 1987), Burrows (1987), and Burrows and Lattimer
(1988) emphasized the importance of neutrinosphere heating by this convection, the latter two
papers pointed out that the transfer of heat to the neutrinosphere from an entropy spike deeper
in the core can accelerate mantle collapse and thereby inhibit the outward progress of the shock
(see also Bethe, Brown, and Cooperstein 1987).

For entropies exceeding $\sim 5$, a negative gradient in $\ln s$ is more destabilizing than
a negative gradient in $\ln Y_{\ell}$ of the same magnitude (\eg, Bruenn, Mezzacappa, and
Dineva 1995, Fig. 1); therefore, one might expect a vigorous entropy-driven protoneutron star
convection to become established exterior to 0.7 - 1.0 \msol\ immediately after shock
propagation. However, the magnitude and location of the negative entropy gradients
established by the dissipating shock depends sensitively on the numerical model used
to compute the prior evolution of the core (Bruenn and Mezzacappa 1994, Fig. 1). The more
realistic the numerical model, the smaller the magnitude of the negative entropy gradients
established by the shock. It is therefore likely that both the magnitude and the extent of the
negative entropy gradients established by the shock will be minimal. Once entropy-driven
protoneutron star convection gets established, the convective mixing will tend to flatten
the entropy gradient driving it; there is no process maintaining the negative entropy
gradient, and the force driving entropy-driven convection will dissipate. Thus, entropy-driven 
protoneutron star convection sows the seeds of its own destruction. A final problem with 
entropy-driven protoneutron star convection for shock revival is that numerical simulations
indicate that conditions favorable for shock revival are not established until $\sim 50$ ms 
after bounce (Janka and M\"{u}ller 1993a; Bruenn and Mezzacappa 1994). On the other hand, 
multidimensional simulations of entropy-driven protoneutron star convection show that the 
convection quickly grows into the nonlinear regime (10 - 15 ms), and the entropy gradients 
are flattened in $\sim$ 25 -- 30 ms (Burrows and Fryxell 1992, 1993; Janka and M\"{u}ller 
1993b, 1996; M\"{u}ller 1993; M\"{u}ller and Janka 1994). Thus, by the time the region above 
the neutrinosphere is in a configuration favorable for neutrino energy deposition, entropy-driven 
protoneutron star convection has dissipated. It is therefore unlikely that it will revive 
the shock. However, it may play a role in seeding both lepton-driven protoneutron star 
convection and neutrino-driven convection. It will also affect the structure of the core, 
and should not be ignored.

\subsection{Current Status of Protoneutron Star Convection}

The role played by entropy- and lepton-driven protoneutron star convection in the supernova
mechanism and the evolution of the core is highly unclear and contradictory at the present
time. Burrows and Fryxell (1992) found vigorous entropy-driven protoneutron star convection
in {\small 2-D} hydrodynamics computations that did not include neutrino transport. With the inclusion
of radial-ray neutrino diffusion they found a ``convective trigger,'' \ie, convectively
enhanced neutrino luminosities that triggered an explosion within 20 ms (Burrows and Fryxell
1993). In a later work (Burrows, Hayes, and Fryxell 1995), they report that entropy- and 
lepton-driven protoneutron star convection are ``weak,'' and presumably unimportant. Herant 
\etal\ (1994), using an SPH hydrodynamics method, found intermittent and model-dependent 
lepton-driven convection in a 
relatively narrow region in the vicinity of the neutrinosphere, and entropy-driven convection 
farther out --- the latter ultimately merging with neutrino-driven convection. Janka and 
M\"{u}ller (1993b, 1996), M\"{u}ller (1993), and M\"{u}ller and Janka (1994) performed 2- 
and 3-D hydrodynamics simulations of protoneutron star convection with several postcollapse 
models and found that the convection, while model dependent, is initially vigorous in all 
cases. In their most recent work they report that convection's purely hydrodynamic effect 
pushes the shock from its stagnation radius at $\sim 200$ km (one-dimensional initial conditions) 
out to $\sim 400$ km (two-dimensional case) and to $\sim 300$ km (three-dimensional case).  Keil, 
Janka, and M\"{u}ller (1996) have recently followed the evolution of a protoneutron star 
for more than 1 s with {\small 2-D} hydrodynamics and radial-ray neutrino diffusion. They found that
lepton-driven Ledoux convection continues for a long time and engulfs the entire core
after about 1 s. Moreover, they found that the convection increases the neutrino luminosities 
by up to a factor of 2 and the mean energies of the emitted neutrinos by 10--20\%. 
Bruenn and Mezzacappa (1994) performed self-consistent {\small 1-D} simulations
of protoneutron star evolution using multigroup flux-limited diffusion (\mgfld) for
neutrino transport and a mixing-length algorithm for convection. In contrast to the
results of Janka and M\"{u}ller, they found only mild convective activity inside and
around the neutrinosphere for a duration $\sim 30$ ms, although there simulations
covered a postbounce period of 0.5 s. While one may argue that the mixing-length
approximation fails to reproduce all of the properties of the convective flow, it 
is nevertheless sensitive to any Ledoux instabilities, and would trace out these
instabilities as they develop.

\vspace*{-0.125in}
\subsection{This Work}

Part of the reason for the present confusion concerning the vigor and extent of
protoneutron star convection and its role in the supernova mechanism is that
numerically simulating this convection is a very difficult problem in radiation
hydrodynamics. Neutrinos and matter are strongly coupled in this regime, and
it can be expected that neutrino transport will have a profound influence on both the
character and the growth rate of any fluid instability. Different initial postcollapse
models and alternative approximations to the neutrino transport can produce widely
varying results. A self-consistent calculation of protoneutron star convection requires
that both the hydrodynamics and the neutrino transport be multidimensional. Only Herant 
\etal\ (1994) have performed such calculations, although their neutrino transport was 
much simpler than the transport used in sophisticated 1-D codes. Helpful insights into the nature 
and growth rates of fluid instabilities in this regime can be obtained by semianalytical 
investigations that are now in progress (Bruenn and Dineva 1997). 
%Past simulations of protoneutron star convection
%have either ignored neutrino transport entirely, or have employed radial-ray neutrino
%transport. While the latter is necessitated by the extreme computational challenge
%imposed by true multidimensional transport, it ignores the tendency of neutrino transport
%to equilibrate upflows and downflows with each other in energy and leptons, thus reducing the
%very buoyancy contrast driving these flows.

In this paper we investigate protoneutron star convection with a numerical scheme
that is a vast improvement over the {\small 1-D} scheme used by Bruenn and Mezzacappa 
(1994). Here we couple {\small 2-D} \ppm\ hydrodynamics with sophisticated {\small 1-D} \mgfld. 
Computations of protoneutron star convection that ignore neutrino transport or 
that implement radial-ray neutrino transport ignore the tendency of transport 
%to reduce the buoyancy contrasts between adjacent radial upflows and downflows
%in the convecting region. 
to equilibrate upflows and downflows with each other in energy and leptons, 
thus reducing the very buoyancy contrast driving these flows.
We show by a simple analytical example that these computations
therefore overestimate the growth rates and final amplitude of protoneutron star 
convection. Our {\small 1-D} neutrino transport represents an alternative intermediate step 
(compared with radial-ray neutrino transport) along the way to full multidimensional 
transport. {\small 1-D} transport obtains in the limit in which the neutrino transport in the 
tangential direction is rapid enough to render the neutrino fields spherically 
symmetric, and therefore, in this regard, is complementary to radial-ray transport. 
It is also important to note that our neutrino transport is multigroup, rather
than gray, which becomes particularly important near the neutrinospheres, where
the matter becomes semitransparent to neutrinos and where gray transport would 
not be adequate. 
%To take multidimensional effects into account, we modify our neutrino interaction 
%rates by introducing a time scale, which is a function of neutrino energy, for the 
%equilibration of separated radial flows that takes into account the time required
%for neutrinos to propagate between convecting fluid elements. 

In Section~\ref{sec-Models}, we describe our hydrodynamics code, ``{\small EVH-1},'' 
our procedure coupling neutrino transport to {\small EVH-1}, and the models used 
to initiate the simulations. In Section~\ref{sec-Results} we present the results 
of our simulations. The tendency of energy and lepton transport by neutrinos to 
suppress protoneutron star convection, which is the central feature in all of our 
numerical results, is discussed with the aid of a simple analytical model in 
Section~\ref{sec-Analytic}. A summary of our work and of its implications is 
given in Section~\ref{sec-Conclusions}.   
\newpage

\section{Initial Models and Methodology}
\label{sec-Models}

\subsection{Overview}

In this work all simulations begin with either
the 15 ${\rm M}_{\odot}$ or the
25 ${\rm M}_{\odot}$ precollapse models, S25s7b or S15s7b, provided
by Woosley (Woosley \& Weaver 1995, Weaver \& Woosley 1997). These 
initial models were evolved through
core collapse and bounce using {\small MGFLD} neutrino transport
and one-dimensional Lagrangian hydrodynamics (Bruenn 1985). These 
1-D simulations were continued approximately 700 milliseconds after 
bounce for the purpose of generating boundary and neutrino distribution 
data to be used in our 2-D simulations, as described below. 
At a simulation time of typically tens
of milliseconds after bounce, the one-dimensional configurations
were mapped onto a two-dimensional Eulerian grid.
The inner and outer boundaries of the two-dimensional grid were
chosen to be at radii
of 20 km and 1000 km, respectively, for all runs. These choices
set the inner and outer boundaries well below and above
the unstable regions at the onset of the simulations.
%Spherically symmetric time-dependent boundary data for the
%two-dimensional hydrodynamics were supplied by the accompanying
%one-dimensional {\small MGFLD} run.
The grids used 128 nonuniform radial
spatial zones. This gave sufficient resolution at the inner
boundary and at the shock. The nonuniform zoning was varied until,
for a test simulation,
excellent agreement was seen between one-dimensional runs
using 128 nonuniform and 512 uniform zones (see Figures 2c--d).
128 uniform angular zones
spanning a range of 180 degrees
%and reflecting boundary conditions
were used for $\theta$. With the assumption of
axisymmetry, the range of 180 degrees spans the entire physical space,
and accordingly, reflecting boundary conditions are used
on the polar axis.

Because the finite differencing in our {\small PPM} scheme is
nearly noise free, and because we cannot rely on machine roundoff
to seed convection in a time that is short compared with
the hydrodynamics time scales in our runs, we seeded
convection in the Ledoux unstable regions below and around
the neutrinospheres by applying random velocity perturbations to
the radial and angular velocities between $\pm$ 5\% of the local
sound speed. This is consistent with seeding used by other groups. 
%An appropriate choice for the 
%magnitudes of these perturbations is still an open question. 
Moreover, in a 2-D simulation of convection during the shell burning of 
oxygen prior to core collapse, Bazan and Arnett (1994) found inhomogeneities of 
density, temperature, pressure, and electron fraction $\sim$ 5\%, 1\%, 
3\%, and 0.08\%, respectively, and convection velocities $\sim$ 15\% 
of the local sound speed. Therefore, our seeding choice seems 
reasonable, although a more systematic study should be carried 
out.
%Using these results as a guide, we chose a 
%magnitude between $\pm$ 5\% of the local sound speed for our 
%velocity seeds.

The two-dimensional hydrodynamics was evolved using {\small EVH-1},
an extended version
of the {\small PPM} hydrodynamics code
{\small VH-1}.
One of the extensions
allows for coupling to general equations of state. For the
most part, matter in our simulations is in nuclear
statistical equilibrium ({\small NSE}), and  to describe its thermodynamic
state we used the equation of state provided by Lattimer and Swesty
(Lattimer \& Swesty 1991). However, below a density and temperature
threshold, the code switches to
a non-{\small NSE} equation of state.
If a zone falls below the density or temperature
threshold of $1.674\times 10^{7}$ \gcm\ or 0.3447 MeV, respectively, it is ``deflashed" 
to silicon, while conserving energy
within the zone. This is well outside the convectively unstable region, 
and deflashing was therefore performed in a spherically symmetric 
manner. An ideal gas equation of state,
plus internal degrees of freedom to mimic the {\small NSE}
equation of state and to provide a more seamless transition to
non-{\small NSE}, is then
used to describe the silicon in its subsequent evolution.
As will be described in detail later,
with both equations of state,
tables of thermodynamic quantities
are constructed
at the beginning of a simulation,
and as the calculation proceeds, the
code interpolates within these tables.

At each time step in the two-dimensional evolution, the results of
the corresponding one-dimensional {\small MGFLD} run, completed at 
a prior time, provide
spherically symmetric time-dependent boundary data for our
inner and outer two-dimensional boundaries,
and neutrino distributions for the local neutrino
heating, cooling, and matter deleptonization
everywhere on our two-dimensional grid.
The boundary data consist of enclosed mass, radial
velocity, density, temperature, electron fraction,
pressure, and specific internal energy, and values at
our fixed inner Eulerian radius were extracted from the
one-dimensional Lagrangian data by interpolation in $r$ and $t$.
A call to the equation of state provides the boundary value
of the first adiabatic index, which is required by our {\small
PPM} scheme.
The outer boundary data were specified in the same way.
The neutrino heating and cooling, and the matter deleptonization,
were computed at each time step using
tables of $\psi^{0}_{i}$'s
in $r$, $t$, and $E_{\nu}$, constructed from our one-dimensional
simulations.
[$\psi^{0}_{i}(r,t,E_{\nu})$ are the zeroth moments of the
neutrino distribution functions; $i$ is a neutrino flavor index.]
Ideally, the neutrino-matter momentum transfer should also be 
included. However, this is a small correction to the matter pressure. 
%We have not included the neutrino-matter momentum transfer in these 
%simulations, but plan to do so in future simulations as we develop 
%more sophisticated multi-D radiation hydrodynamics. 

Gravity was assumed to be Newtonian and spherically symmetric
in our two-dimensional simulations.
The justification for the latter is the fact that the gravitational field in the
convectively unstable region was dominated by the spherically symmetric enclosed mass
at the region's base.
The time dependence of
the mass enclosed by our inner boundary, given by our
one-dimensional {\small MGFLD} runs, was taken into account.
The contribution to the gravitational field
of the mass on the grid was calculated
at each time step by finding the angle--averaged density
in a shell between radial grid points, and summing the masses
of the shells up to a particular radial grid point.
(The solution of the Poisson equation for the
gravitational potential will be incorporated in future investigations.)

Comparisons were made (see below) to ensure that in one
dimension the results from our code matched the results obtained
with the {\small MGFLD} code, modulo {\small EVH-1}'s better
resolution of the shock.

\vspace*{-0.375in}
\subsection{Details of the Simulations}

Table 1 contains the initial times for, and
a brief description of, each simulation. All of the simulations
listed below were performed with the {\small PPM} code, and began
a few milliseconds after bounce. The one-dimensional {\small PPM}
simulations were performed for the purpose of comparison with our
one-dimensional {\small MGFLD} simulations.

In simulations B, D, J, and L, the electron neutrino and
electron antineutrino distributions ($\psi^{0}_{\nu_{\rm e},
\bar{\nu}_{\rm e}}$)
used to compute the local neutrino heating, cooling, and
matter deleptonization are generated by our accompanying one-dimensional
simulations that include all neutrino flavors and
interactions --- neutrino--electron scattering (NES) and
ion--ion screening corrections included.

Simulations F, G, and H were initiated from a model whose prior 
evolution involved a less rigorous treatment of neutrino 
transport, and were dubbed our ``low-test" runs. In particular, the initial 
model for these runs was evolved through bounce by the {\small MGFLD}
code, neglecting NES and ion--ion screening corrections.
This less rigorous treatment of neutrino transport
provides initial conditions for the two-dimensional
simulations that are more conducive to the development of
protoneutron star convection because
a larger entropy gradient is established by the weakening of
an originally stronger shock.

To explain the way the neutrino distributions are
used to calculate the local neutrino heating, cooling, and
deleptonization, we must describe how {\small EVH-1} carries out the
simulations:

{\small EVH-1} performs the
hydrodynamics evolution in sweeps over the grids.
In the one-dimensional
case, there is a sweep in the radial direction. In the
two-dimensional case, there is a sweep first in the radial
direction for each angle and then in the angular direction
for each radius.
In a particular
sweep, there are four principal variables that evolve. These
are density ($\rho$), temperature ($T$), electron fraction
($Y_{\rm e}$), and specific internal energy ($\epsilon$).
($Y_{\rm e}$ in a zone changes due to hydrodynamic advection,
in addition to transformations between $e^{-}$'s, $e^{+}$'s,
$\nu_{{\rm e}}$'s, and $\bar{\nu}_{{\rm e}}$'s.)
In order to update these quantities, thermodynamic quantities such
as the pressure are required. In order to calculate these
quantities,
{\small EVH-1} calculates tables of state variables
and interpolates within these tables. The three-dimensional
tables are constructed
such that $\rho$, $T$, and $Y_{\rm e}$ are the independent variables,
and tables are constructed for $\epsilon$,
pressure ($P$), entropy ($s$), and other variables.
For the combination $\rho$, $T$, and $Y_{\rm e}$
in a given zone of our two-dimensional grid,
the code performs linear interpolations using the logarithms of
both the independent (\eg, $\rho$) and the dependent (\eg, $P$) 
variables in our table to obtain the corresponding dependent variables for that zone.
For details of the interpolation scheme,
see Mezzacappa and Bruenn (1993).

Because
the hydrodynamics evolution updates $\rho$, $\epsilon$, and $Y_{\rm e}$,
it is necessary to update $T$ by iteration of

\begin{equation}
\Delta T \: = \: \frac{   \epsilon^{\rm hyd} -
\epsilon^{\rm int}   }
{  \left( \frac{ \partial \epsilon}{\partial T} \right)_{\rho,Y_{\rm e}}}
\; \: ,
\label{delt}
\end{equation}
\smallskip

\noindent where $\epsilon^{\rm hyd}$ and $\epsilon^{\rm int}$
are the specific internal energy (a) partially updated
by the hydrodynamics and (b) given by
the interpolation, respectively; and
${  \left( \frac{ \partial \epsilon}{\partial T} \right)_{\rho,Y_{\rm e}}}$
is the specific heat.
$\epsilon^{\rm int}$ and
${  \left( \frac{ \partial \epsilon}{\partial T} \right)_{\rho,Y_{\rm e}}}$
are evaluated at the current iterate, $T^{0}$.
The iterations occur until $\Delta T$ falls
below a specified tolerance.

The sweep proceeds by first calling the
interpolation routine to update $T$, and then calculates variables such as
$P$ and $s$ in each zone of a one-dimensional sweep.
Next, the sweep proceeds by
calling the {\small PPM} routines for the hydrodynamics
evolution, partially updating $\rho$, $\epsilon$,
and $Y_{\rm e}$.
Then, the simulations make
use of the {\small MGFLD} neutrino data in an operator split
manner to fully update $\epsilon$ and $Y_{\rm e}$
according to the
prescriptions below. The next sweep repeats this procedure,
first updating
$T$, then evolving the hydrodynamics, and further updating
$\epsilon$ and $Y_{\rm e}$ due to neutrino emission and
absorption.

In simulations B, D, F, and L,
the neutrino heating and cooling, and the change in the electron fraction,
were computed at each time step within each angular sweep (i.e.,
at a given radius) using the following formulae:

\begin{equation}
\partial\epsilon /\partial t=c\sum_{i=1}^{2}\int E_{\nu}^{3}dE_{\nu}
[\psi^{0}_{i}/\lambda^{(a)}_{i}-j_{i}(1-\psi^{0}_{i})]/\rho (hc)^{3},
\label{epsdot}
\end{equation}
\smallskip

\noindent and

\begin{equation}
\partial Y_{\rm e}/\partial t=cm_{\rm B}\sum_{i=1}^{2}\alpha_{i}\int E_{\nu}^{2}dE_{\nu}
[\psi^{0}_{i}/\lambda^{(a)}_{i}-j_{i}(1-\psi^{0}_{i})]/\rho (hc)^{3},
\label{yedot}
\end{equation}
\smallskip

\noindent where
$E_{\nu}$, $\psi^{0}_{i}$, $\lambda^{(a)}_{i}$, and $j_{i}$ are the
electron neutrino or antineutrino energy, zeroth distribution function
moment, absorption mean free path, and emissivity, respectively;
$m_{\rm B}$ is the baryon mass; $i=1,\alpha_{1}=1$ corresponds to
electron neutrinos, and $i=2,\alpha_{2}=-1$ corresponds to electron
antineutrinos. As mentioned earlier, the time-dependent $\psi^{0}_{i}$'s
are obtained from tables in $r$, $t$, and $E_{\nu}$
constructed from our one-dimensional
{\small MGFLD} simulations.

Simulations E and H made use of modified versions of the
above formulae to account for the ``finite time" required for
transport of neutrinos between an element of convecting material
and its background. The corrected heating rate is calculated
from the above heating rate by considering the mean
free paths and by calculating a minimum transport time, which
is then used to calculate a correction factor. The mean
free path, $\lambda(r, E_{\nu}, i)$, is calculated from
the emission, absorption, and isoenergetic scattering mean
free paths for each radius ($r$), neutrino
energy group ($E_{\nu}$), and for each neutrino type ($i$).
Once $\lambda(r, E_{\nu}, i)$ is determined, the diffusion time,
$t_{\mbox {\scriptsize diff}} (r, E_{\nu}, i)$,
is calculated by

\begin{equation}
t_{\mbox {\scriptsize diff}} \left(r, E_{\nu}, i\right) \: = \:
\frac{\left(r_{\mbox{\scriptsize scale}} \right)^2}{c \lambda(r, E_{\nu}, i)}
\end{equation}
\smallskip

\noindent where $r_{\mbox{\scriptsize scale}}$ is the
length scale of a convecting element. In these calculations, this
scale was taken to be 10 km, which is about one pressure scale
height in the convectively unstable region of the core, 10 to 100 ms
after bounce. The free escape time,
$t_{\mbox{\scriptsize esc}} \left(r, E_{\nu}, i\right)$,
represents
the minimum transport time of neutrinos from the convecting
element, and is given by

\begin{equation}
t_{\mbox{\scriptsize esc}} (r, E_{\nu}, i) \: = \:
\frac{r_{\mbox{\scriptsize scale}}}{c} \; .
\end{equation}
\smallskip

\noindent The free escape time is a physical limit, and it must be used
when $t_{\mbox{\scriptsize diff}} (r, E_{\nu}, i)$ becomes so small,
i.e.,  $\lambda(r, E_{\nu}, i)$ becomes so large,
that it would imply that neutrinos are propagating
faster than the speed of light. The quantity
$t_{\mbox{\scriptsize trans}} (r, E_{\nu}, i)$ smoothly interpolates
between $t_{\mbox{\scriptsize diff}}(r, E_{\nu}, i)$
[small $\lambda(r, E_{\nu}, i)$] and $t_{\mbox{\scriptsize esc}}
(r, E_{\nu}, i)$
[large $\lambda(r, E_{\nu}, i)$], and is given by

\begin{equation}
t_{\mbox{\scriptsize trans}} (r, E_{\nu}, i) \: = \:
\left[ 
  \left( t_{\mbox{\scriptsize diff}}(r, E_{\nu}, i)\right)^2
+ \left( t_{\mbox{\scriptsize esc}} (r, E_{\nu}, i)\right)^2
\right]^{1/2} \: \; .
\label{ttrans}
\end{equation}
\smallskip

\noindent This expression tends to weight preferentially the larger of
$t_{\mbox{\scriptsize diff}} (r, E_{\nu}, i)$ and $t_{\mbox{\scriptsize esc}}
(r, E_{\nu}, i)$.
Next, from $[\partial\epsilon /\partial t] (r, E_{\nu}, i)$, which is given by
the individual terms in the integrand of equation (\ref{epsdot}),
the heating time, $t_{\mbox{\scriptsize e}} (r, E_{\nu}, i)$,
is given by

\begin{equation}
t_{\mbox{\scriptsize e}} (r, E_{\nu}, i) \: = \:
\frac{\epsilon (r)}{\left[\partial\epsilon /\partial t\right] (r, E_{\nu}, i)} \; .
\label{ctrans}
\end{equation}
\smallskip

\noindent The correction factor, $c_{\mbox{\scriptsize trans}}$, is then

\begin{equation}
c_{\mbox{\scriptsize trans}} (r, E_{\nu}, i) \: = \:
\frac{ t_{\mbox{\scriptsize e}} (r, E_{\nu}, i) } {
t_{\mbox{\scriptsize e}} (r, E_{\nu}, i) + t_{\mbox{\scriptsize trans}}
(r, E_{\nu}, i) } \; \: ,
\label{cordedt}
\end{equation}
\smallskip

\noindent and the corrected heating rate becomes

\begin{equation}
\left(\partial\epsilon /\partial t\right)_{\mbox{\scriptsize corrected}} \: = \:
c_{\mbox{\scriptsize trans}} (r, E_{\nu}, i) \partial\epsilon /\partial t
\; .
\label{corrate}
\end{equation}
\smallskip

\noindent A correction for the deleptonization rate is 
computed and applied in an analogous manner. 

\subsection{Initial Conditions}

Table 2 contains the initial conditions for our simulations. The unstable region 
in the 25 ${\rm M}_{\odot}$ model, from which our ``25 ${\rm M}_{\odot}$'' simulations 
are initiated, extended from an enclosed mass of 0.82 M$_{\odot}$ to an enclosed mass 
of 1.02 M$_{\odot}$ for the case in which the prior evolution (up to postbounce) was 
calculated with comprehensive neutrino transport. For the ``low-test" case, the 
region extended from an enclosed mass of 0.97 M$_{\odot}$ to an enclosed mass of
1.21 M$_{\odot}$. For the 15 ${\rm M}_{\odot}$ model, from which our ``15 ${\rm M}_{\odot}$'' 
simulations are initiated, the unstable region extended from an enclosed mass of
0.79 M$_{\odot}$ to an enclosed mass of 1.03 M$_{\odot}$.
\newpage

\section{\bf Results}
\label{sec-Results}

In Figure 1, we plot the initial entropy and electron fraction
gradients for our 15 and 25 \msol\  models. In the 25 \msol\
case,  we plot the gradients for both our original and ``low-test''
runs. Note that in the latter case there is a significantly larger 
initial entropy gradient. ``Low-test'' transport during core collapse 
leads to less deleptonization, and consequently, to a larger inner 
homologously collapsing core. As a result, at bounce a stronger 
shock forms (from a larger rebounding inner piston), and it forms 
farther out in radius. All are evident in the figure. For our 
original 25 \msol\ model, the neutrinospheres are initially 
located at (a) 71 km, (b) 63 km, and (c) 73 km for (a) electron 
neutrinos, (b) electron antineutrinos, and (c) muon and 
tau neutrinos and antineutrinos. For our ``low-test'' 
runs, the neutrinospheres are initially located at (a) 117 km, 
(b) 110 km, and (c) 119 km for (a) electron neutrinos, 
(b) electron antineutrinos, and (c) muon and tau 
neutrinos and antineutrinos. For our 
15 \msol\ model, the neutrinospheres are initially 
located at (a) 68 km, (b) 59 km, and (c) 69 km for (a) electron 
neutrinos, (b) electron antineutrinos, and (c) muon and 
tau neutrinos and antineutrinos.

As a check that our inner and outer running boundary 
conditions and our procedure for computing the neutrino 
cooling, heating, and matter deleptonization rates are 
sufficient to ensure that {\small 1-D} {\small EVH-1} simulations between the 
inner and outer boundary closely match the original 
\mgfld\ simulations, 
%As a check on our simulations, in Figure 2a and 2b we plot a
in Figure 2a and 2b we plot a
comparison of the density, entropy, electron fraction, and
velocity profiles for the one-dimensional simulation, B, and
the corresponding \mgfld\ simulation. These two simulations were initiated from the 
same model and then run independently thereafter (except, of 
course, that simulation B uses the neutrino and boundary data 
generated by the \mgfld\ run). The comparison is made at 71 ms
after bounce. 
Modulo differences at the shock, the agreement between the two simulations
is excellent. Differences are expected at the shock because
the \ppm\ Godunov-type scheme is better able to resolve it
than second-order artificial viscosity methods.
%(1) the \ppm\ Godunov-type scheme is better able to resolve it
%than second-order artificial visocity methods, and (2) 
%even more extensive rezoning (than the already extensive 
%rezoning performed) would be required in the Lagrangian 
%hydrodynamics code to maintain the zoning resolution in the region 
%of the shock that is present in the {\small EVH-1} grid.

In Figure 3a, we include three two-dimensional entropy slices 
spanning the first 20 ms of our simulation C, which is a 
``hydrodynamics only'' model. The development of protoneutron star 
%convection in the innermost region below the neutrinospheres 
convection in the innermost region below the neutrinospheres 
(34 -- 52 km)  
%is evident. In the absence of neutrino transport, the entropy 
%and electron fraction gradients in this initial Ledoux unstable 
%region are smoothed out by convection, and the convection ceases. 
%The smoothing is shown in Figure 3c, where we plot the angle-averaged
%entropy and electron fraction versus radius at the same three 
%time slices. These quantities are defined as follows
is evident. In the absence of neutrino transport, the redistribution
of entropy and electron fraction by convection stabilizes this region,
and convection dies out by $t_{\rm pb} = 28$ ms. The
redistribution of the entropy and electron fraction profiles by
convection is shown in Figure 3c, where we plot the angle-averaged
entropy and electron fraction versus radius at the same three
time slices shown in Figure 3a. These quantities are defined as follows 

\begin{equation}
< S >(i) = \frac{1}{A(i)} \sum^{n_{\theta}}_{j = 1} A(i,j) S(i,j)
\label{eq:avgent}
\end{equation}
\smallskip

\noindent and

\begin{equation}
< Y_{\rm e} >(i) = \frac{1}{A(i)} \sum^{n_{\theta}}_{j = 1} A(i,j) Y_{\rm e}(i,j)
\label{eq:avgye}
\end{equation}
\smallskip

\noindent where

\begin{equation}
A(i,j)=2\pi r^{2}(i)\sin\theta(j)d\theta
\label{eq:area}
\end{equation}
\smallskip

\noindent and where $A(i)=4\pi r^{2}(i)$ and $d\theta =\pi/128$.
In equations (\ref{eq:avgent}) and (\ref{eq:avgye}), $S(i,j)$
and $Y_{\rm e}(i,j)$ are our two-dimensional entropy
and electron fraction configurations; the index $i$ runs over
$n_{r}=128$ radial zones, and the index $j$ runs over $n_{\theta}=128$
angles. As $\theta\longrightarrow 0$, the area of
the strip on the 2-sphere spanned by $d\theta$, $A(i,j)$, approaches zero. 
The angle averaging in equations (\ref{eq:avgent}) and (\ref{eq:avgye})
is therefore designed to ensure that the averaged quantities 
near the poles contribute less. In Figure 3c, it is seen that at $t_{\rm pb} = 13$ ms
both entropy and electron fraction gradients are negative below 40 km,
but by $t_{\rm pb} = 18$ ms, the entropy gradient has become positive and 
is thus tending to stabilize the still negative gradient in 
electron fraction. 

The additional entropy gradients in Figure 3c (\eg, at 100 km) that appear below 
the shock with time result from the successive strengthening 
and weakening of the shock as it first propagates out after bounce 
and as the inner core oscillates in
quasihydrostatic equilibrium. These gradients give rise to
additional Ledoux unstable regions, and convection develops
in these regions, seeded by the initial convection episode.
This is evident in Figure 3a, at $t_{\rm pb} = 28$ ms. The shock is clearly outlined by 
the large jump in entropy across it. The distortions in the 
shock's sphericity when protoneutron star convection reaches it are 
also evident. 

In Figure 3b, we plot the angle-averaged radial and angular 
convection velocities, defined by

\begin{equation}
< v_{c} >_{r} = \frac{1}{A(i)} \sum^{n_{\theta}}_{j = 1} A(i,j) \mid \: \mid v_r(i,j) \mid - < v > \: \mid
\label{eq:avgrvel}
\end{equation}
\smallskip

\noindent and

\begin{equation}
< v_{c} >_{\theta} = \frac{1}{A(i)} \sum^{n_{\theta}}_{j = 1} A(i,j) \mid v_\theta(i,j) \mid
\label{eq:avgthetvel}
\end{equation}
\smallskip

\noindent where

\begin{equation}
< v > \; = \; \frac{1}{A(i)} \sum^{n_{\theta}}_{j = 1} A(i,j) \mid v_r(i,j) \mid
\label{eq:avgvel}
\end{equation}
\smallskip

\noindent Within 10 ms, the radial convection velocity in the innermost
region peaks between 3.7 and 5.0 $\times 10^{8}$ cm/sec, whereas the 
angular convection velocity peaks at a somewhat larger value of 6.2 
$\times 10^{8}$ cm/sec. The radial convection velocity below the shock 
reaches a peak value of $10^{9}$ cm/sec after about 20 ms; by then, the 
angular convection velocity has decreased. Note the anticorrelation
between the radial and angular convection velocities. The peaks in
the radial velocity correspond to troughs in the angular velocity
and vice versa, which is characteristic of convective ``rollover.'' 

In Figure 4a, we include three entropy slices spanning
21 ms in our simulation D. With the inclusion of neutrino
transport, there is no evidence of protoneutron star convection. 
Providing more quantitative detail, Figure 4b shows 
that the angle-averaged radial and angular convection 
velocities peak at values orders of magnitude smaller 
than in simulation C. In the next section, we will 
present a simple analytical model that illustrates why 
this occurs. 

Very important to the analysis we will present here and in the next
section, and to the overall conclusions of this paper, is the angle-averaged
radial fluid velocity (as opposed to the angle-averaged convection 
velocities). In Figure 4d, early in simulations D and L, we plot the
fluid velocity as a function of radius with a focus on the innermost 
100 km, where protoneutron star convection first develops in both simulations. 
Most important to note is that in simulation C, the convection velocities 
in this region are comparable to the fluid velocity; consequently, 
the convective transport of entropy and leptons is evident in Figure 
3a. On the other hand, we see that the convection velocities in 
simulation D are orders of magnitude smaller than the fluid velocity,
which explains why convective transport of entropy and leptons is not 
evident in Figure 4a.

In addition to the effect on the convection velocities, 
Figure 4c demonstrates transport's effect on the innermost 
entropy gradient in the initial Ledoux unstable region. In 
this case, the entropy gradient is smoothed out by neutrino 
diffusion [see also Burrows \etal\ (1995)], whereas in simulation 
C, it was smoothed out by convection. Moreover, Figure 4c shows 
that the initial lepton gradient is more or less maintained 
by the same neutrino diffusion. More important, despite the 
maintenance of this gradient, we see no evidence of protoneutron 
star convection over the entire 100 ms spanned by our simulation, 
counter to what was seen by Burrows (1987), Burrows and Fryxell 
(1993), Burrows \etal\ (1995), and by Janka and M\"{u}ller 
(1993b, 1994, 1995).

In a self-consistent multidimensional simulation with both
multidimensional neutrino transport and hydrodynamics, matter 
irregularities would give rise to irregularities in the neutrino 
radiation fields. These would be smoothed out by neutrino transport, 
which in turn, coming full circle, would smooth out the original 
matter irregularities. The irregularities in the radiation fields 
would be smoothed out on transport time scales, which we neglect 
by imposing already spherically symmetric neutrino distributions.
To correct for this, in simulation E we take into account the
time it takes for entropy and lepton transport to occur between
a convecting fluid element and its surroundings, by modifying the 
entropy and lepton equilibration rates. This is done as outlined 
in the previous section. Of course, in a fully self-consistent
simulation, the smoothed radiation field may ultimately be 
different than the spherically symmetric fields we impose.
This is a shortcoming of our approximation that we cannot 
correct. However, we find that transport suppresses protoneutron 
star convection significantly, and in so doing, tends to keep 
the matter spherically symmetric, which in turn would keep 
the radiation fields close to our imposed fields. 

%Because we are imposing a background spherically symmetric
%neutrino distribution in our two-dimensional simulations,
%in optically thick regions we neglect (1) the effect trapped
%neutrinos carried with a convecting fluid element have on the
%equlibration of the element with its surroundings and (2) the
%finite transport time required for neutrinos to diffuse into
%or out of that element. 
%Moreover, the imposition of a spherically
%symmetric neutrino distribution is equivalent to maximizing the
%lateral angular transport of neutrinos between ascending
%high-entropy and descending low-entropy convecting fluid 
%elements. Lateral transport of neutrinos, and its corresponding
%expected effect on reducing protoneutron star convection, will require a
%self-consistent coupling of two-dimensional \mgfld\ transport and
%two-dimensional hydrodynamics, which is beyond the scope of this
%work. 
%However, in simulation E, we do take into account the 
%time it takes for entropy and lepton transport to occur between 
%a convecting fluid element and its surroundings. This is done,
%as outlined in the previous section, by modifying the entropy
%and lepton equilibration rates.
%
Figure 5a includes three entropy slices spanning 20 ms in simulation E. 
Again, as in simulation D, where ``finite-time'' effects were not considered,
there is no evidence of protoneutron star convection. More quantitatively, Figure 5b 
illustrates that the peak angle-averaged radial and angular convection 
velocities are almost an order of magnitude larger than they are in 
simulation D, but still roughly two orders of magnitude smaller than
they are in simulation C (a ``hydrodynamics only'' model). Therefore, 
our conclusion that neutrino transport renders 
protoneutron star convection insignificant 
is not changed. Figure 5c shows once again the smoothing of the innermost 
entropy gradient and the maintenance of the lepton gradient by neutrino 
diffusion during the course of the simulation.

To further enhance protoneutron star convection's chances of developing
in the presence of neutrino transport, and to further explore
the interplay between neutrino transport and protoneutron star convection,
we have carried out several runs with ``low-test'' initial conditions, 
in which the initial entropy gradient, which initially drives protoneutron star 
convection, is maximized. Figure 6a includes three entropy slices 
from simulation F, in which neutrino transport is turned off. 
The large initial entropy gradient is evident 
in the first slice, and protoneutron star convection develops rather dramatically 
within 10 ms after the start of our run. After an additional 10 ms (at $t_{\rm pb} = 27$ ms), 
convection has reached the shock and has distorted it significantly. 
Figure 6b shows the peak angle-averaged radial and angular convection 
velocities. At $t_{\rm pb}\sim 16$ ms, the peak convection velocities for 
simulations C and F are comparable, but by $t_{\rm pb}\sim 26$ ms, the 
velocities in the innermost region in simulation F are roughly 
two times larger, and the outer peak in radial convection velocity 
is still developing, indicating that the convection growth time 
for simulation F is longer. Figure 6c shows the rather dramatic 
and rapid smoothing of the initial entropy and lepton fraction 
gradients below $\sim 100$ km that occurs in this simulation.

In Figure 7a, the three entropy slices show minimal protoneutron star convection 
developing within 10 ms of the start of simulation G, which includes 
neutrino transport, but begins with the same ``low-test'' initial conditions 
used in simulation F. However, by $t_{\rm pb}=27$ ms, at least in the 
two-dimensional entropy plot, the configuration appears once
again to be spherically symmetric. Examination of the convection
velocities in Figure 7b shows that, relative to simulation
F, the velocities are reduced, particularly at $t_{\rm pb}=27$ ms,
where they are significantly reduced. Figure 7c shows the
smoothing of the entropy gradient in the initial Ledoux 
unstable region and the maintenance of the lepton gradient 
that is consistently exhibited by all of our two-dimensional 
simulations that include neutrino transport (simulations
D,E,G, and H).
  
To help protoneutron star convection even more, we carried out simulation
H, which uses the ``low-test'' initial conditions used in
simulations F and G, and the reduced entropy and lepton
equilibration rates used in simulation E. Figure 8a shows
three entropy slices from this simulation. Again, at $t_{\rm pb}=17$ 
ms, minimal protoneutron star convection is evident, but after an additional 
10 ms, the two-dimensional entropy configuration appears to
be spherically symmetric. A look at Figure 8b shows that 
the peak radial and angular convection velocities are 5--10 
times smaller relative to their values in simulation F 
(``low-test'' initial conditions; ``hydrodynamics only''
model). Figure 8c shows the evolution in the entropy and 
electron fraction gradients, with the expected smoothing
of the initial entropy gradient during the first 20 ms by 
neutrino diffusion, and the maintenance of the lepton 
gradient.

Figure 9 illustrates the long-term behavior in entropy 
in simulations D, E, and G. No convection is evident
in simulation D at 86 ms into our run, and minimal 
convection is evident in simulation E at about the 
same time. In contrast, in the ``low-test'' simulation, 
G, more vigorous protoneutron star convection has developed at
this later time in our run than appeared at earlier 
times (see Figure 7a).

Figures 10a--c and 11a--c depict the evolution in entropy,
convection velocities, and entropy and electron fraction
gradients for our 15 \msol\ model. The same behavior exhibited
by the 25 \msol\ model is exhibited here, indicating that our 
conclusions regarding neutrino transport's tendency to inhibit the development
of protoneutron star convection are independent of stellar mass 
and the size and profile of the initial precollapse or postbounce 
cores.
\newpage

\section{Analytical Model}
\label{sec-Analytic}

Convection near or below the neutrinospheres can be profoundly 
influenced by the neutrino transport of energy and leptons between 
a convecting fluid element and the background. In effect, convection 
becomes ``leaky,'' and differences between a convecting fluid 
element's entropy and lepton fraction and the background's entropy
and lepton fraction, from which the buoyancy force driving convection 
arises, are reduced. 
To construct the simplest model of this, we will assume that the 
lepton fraction gradient is zero and that convection is driven 
by a negative entropy gradient that is constant in space and time. 
[Reversing the roles of the entropy and lepton fraction gradients 
would give analogous results. The more general case in which both 
gradients are nonzero has been considered 
by Bruenn and Dineva (1996). This complicates the analysis and can 
lead to additional modes of instability, such as semiconvection 
and neutron fingers, which are not relevant here.] We will also
assume that the effect of neutrino transport is to equilibrate 
the entropy of a fluid element with the background entropy in a 
characteristic time scale $\tau_{s}$.\footnote{We realize that 
temperature is the appropriate thermodynamic variable to describe 
the equilibration in energy of two systems in thermal contact, as 
is the chemical potential for equilibration in particle number. Because 
our system is in nuclear statistical equilibrium, it is legitimate 
to consider equilibration in terms of the variables $s$ and $Y_{\ell}$; 
\ie\ , when the temperatures and chemical potentials of the two 
systems in pressure balance become equal, their $s$'s and 
$Y_{\ell}$'s are also equal. Using the same variables to describe equilibration 
that are used to describe convective instability greatly simplifies 
the picture.} If the fluid element and 
the background are in pressure balance, and if we neglect viscosity 
(microscopic and turbulent) and keep only first order terms, the fluid element's 
equations of motion are 

\vspace*{-0.125in}
\begin{equation}
\dot{v} = \frac{g}{\rho} \alpha_{s} \theta_{s}
\label{eq:zddot} 
\end{equation}
\smallskip

\begin{equation}
\dot{\theta}_{s} = - \frac{\theta_{s}}{\tau_{s}} - \frac{d\bar{s}}{dr} v
\label{eq:thetadot} 
\end{equation}
\smallskip

\noindent where $\theta_{s} = s - \bar{s}$, with $s$ and $\bar{s}$ being 
the fluid element's entropy and the background's entropy, respectively; $g$ is 
the local acceleration of gravity; $v$ is the radial velocity; and 
$\alpha_{s} \equiv - \left( \partial \rho /\partial s 
\right)_{P,Y_{\ell}} > 0$, where $Y_{\ell}$ is the common lepton 
fraction. Equation (\ref{eq:zddot}) equates the 
fluid element's acceleration to the buoyancy force arising from 
the difference between its entropy and the background's entropy; equation 
(\ref{eq:thetadot}) equates $\dot{\theta}_{s}$ to $\dot{s}$ minus 
$\dot{\bar{s}}$, where $\dot{s}$ results from the fluid element's 
equilibration with the background, and $\dot{\bar{s}}$ results from 
its motion through the gradient in $\bar{s}$. 

If we neglect neutrino effects ($\tau_{s} = \infty$), the solutions
to equations (\ref{eq:zddot}) and (\ref{eq:thetadot}) indicate that
(a) if $d\bar{s}/dr > 0$, the fluid element oscillates with the 
Brunt-V\"{a}is\"{a}l\"{a} frequency $\omega_{BV_{s}} \equiv \left[ 
(g\alpha_{s}d\bar{s}/dr)/\rho \right]^{1/2}$, and (b) if $d\bar{s}/dr 
< 0$, it convects, i.e., its velocity increases exponentially, and the 
convection growth time scale is given by $\tau = \tau_{BV_{s}} \equiv 
\left[ -(g\alpha_{s}d\bar{s}/dr)/\rho \right]^{-1/2}$. 
When neutrino transport effects are included in the convectively 
unstable case ($d\bar{s}/dr < 0$), the fluid element convects,
but the convection growth time scale $\tau > \tau_{BV_{s}}$ is given by 
$1/\tau = \left[ 1/\tau_{BV_{s}}^{2} + 1/4\tau_{s}^{2} \right]^{1/2}- 1/2\tau_{s}$. 
In the limit $\tau_{s} \ll \tau_{BV_{s}}$, the growth time scale increases by 
$\tau_{BV_{s}}/\tau_{s}$, i.e., $\tau \simeq \tau_{BV_{s}}^{2}/\tau_{s}$. 

In addition to reducing convection's growth rate, neutrino transport 
also reduces its asymptotic velocities. In particular, in the limit 
$\tau_{s} \ll \tau_{BV_{s}}$, the solutions to equations (\ref{eq:zddot}) 
and (\ref{eq:thetadot}) show that a fluid element's velocity after 
moving a distance $\ell$ from rest is reduced by the factor $\tau_{s}/\tau_{BV_{s}}$. 

To apply this analysis to protoneutron star convection, we note that $\tau_{BV_{s}}
=1.8$ ms at the neutrinosphere ($\rho\sim 3\times 10^{11}$ g/cm$^{3}$) and 1.4
ms at $10^{12}$ g/cm$^{3}$. These values are representative of both models, S15s7b 
and S25s7b, after bounce, and are computed for a typical postbounce $s$
gradient of 1, in 30 km, and no gradient in $Y_{\ell}$. Similarly,  $\tau_{BV_{Y_{\ell}}}
=12.3$ ms at the neutrinosphere and 16.6 ms at $10^{12}$ g/cm$^{3}$. 
These values are computed for a postbounce gradient of 0.1, in 30 km, and 
no gradient in $s$.
On the other hand, our $\dot{\epsilon}$ from neutrino heating and cooling, including
``finite-time'' corrections, gives values for $\tau_{s}$ that decrease from 0.62 ms 
at the neutrinosphere to 0.03 ms at $10^{12}$ g cm$^{-3}$, for our 25 \msol\ model,
and from 0.65 ms at the neutrinosphere to 0.028 ms at $10^{12}$ g cm$^{-3}$, for our 15 
\msol\ model. Note that the numbers are fairly independent of the precollapse model. 
Our $\dot{Y}_{e}$, with ``finite-time'' corrections, gives lepton equilibration times 
of 0.049 ms at the neutrinosphere and 0.018 ms at $10^{12}$ g cm$^{-3}$, for our 25 
\msol\ model, and 0.054 ms at the neutrinosphere and 0.017 ms at $10^{12}$ g cm$^{-3}$, 
for our 15 \msol\ model. 
These imply that neutrino transport should reduce the growth rate and asymptotic velocities 
of entropy-driven (lepton-driven) convection by a factor $\sim$ 3 (50) at the neutrinosphere 
and a factor $\sim 250$ (1000) at $10^{12}$ g cm$^{-3}$, for both the 15 and 25 \msol\ models. 
Our simulations are consistent with these results. 
\newpage

\section{\bf Summary, Discussion, and Conclusions}
\label{sec-Conclusions}

In the absence of neutrino transport, for both our 15 and 25 \msol\ models 
the angle-averaged radial and angular convection velocities in the
initial Ledoux unstable region below the shock after bounce achieve
their peak values in $\sim$ 20 ms, after which they decrease as the
convection in this region dissipates. The dissipation occurs as the initial 
postbounce entropy and lepton gradients are smoothed out by convection. 
The initial protoneutron star convection episode seeds additional 
protoneutron star convection beneath the shock, which is driven by 
successive entropy gradients that develop above the initially unstable 
region, as the shock propagates out after core bounce and is in turn 
strengthened and weakened by the oscillating inner core.

In the presence of neutrino transport, protoneutron star convection velocities are
too small relative to the bulk inflow velocities to result in any significant
convective transport of entropy and leptons. Our two-dimensional entropy 
profiles remain for all intents and purposes spherically symmetric, and
the peak angle-averaged radial and angular convection velocities are orders 
of magnitude smaller than they are in the corresponding ``hydrodynamics only'' 
models. 

When neutrino transport is included in any of our simulations, the initial 
postbounce entropy gradient is smoothed out by neutrino diffusion, whereas 
the initial lepton gradient is maintained by electron capture and neutrino 
escape near the neutrinosphere. Counter to what was seen by Burrows (1987), 
Burrows and Fryxell (1993), Burrows \etal\ (1995), and by Janka and M\"{u}ller 
(1993b, 1994, 1995), despite the persistence of the lepton gradient, protoneutron 
star convection develops in the 100 ms spanned by our simulations only in our 
``low-test'' simulation, which begins with unrealistic initial entropy and 
lepton gradients that result from an unrealistic core collapse simulation with 
oversimplified neutrino transport. 

%In optically thick regions, the use of a precalculated spherically symmetric 
%neutrino distribution in our two-dimensional simulations neglects (1) the 
%effect trapped neutrinos have on the equlibration of a convecting fluid 
%element with its surroundings and (2) the finite time required for neutrino 
%diffusion between the two. The two factors are actually intertwined. For 
%%neutrino distribution in our two-dimensional simulations neglects  
%%the finite time required for neutrinos to diffuse between the convecting 
%%fluid elements and the background, and therefore overestimates the 
%%equilibration rates between the two. For
%model S15s7b, the Planck-averaged optical depth from the top to the bottom 
%of the initial unstable region varies from 2.0 to 31 for electron neutrinos and from 
%0.8 to 16.5 for electron antineutrinos. The corresponding quantities for 
%S25s7b vary from 2.9 to 15 and from 1.1 to 5.3, respectively. Therefore, 
%at least at the base of our initial Ledoux unstable regions, we expect to
%seriously overestimate the neutrino equilibration rates.
%
%To correct for this shortcoming, 
We have included simulations in 
which the entropy and lepton equilibration rates (as defined in 
Section 2) were reduced by considering the neutrino diffusion 
time across a convecting element of pressure-scale-height size; 
i.e., by using the largest convecting element, we overestimate the
equilibration suppression. Details can be found in Section 2. 
Despite the reduction in rates, our conclusions are not altered. 
The two-dimensional entropy profiles show minimal protoneutron star convection 
development, and the ratio of the angle-averaged radial and angular 
convection velocities to their counterparts in the ``hydrodynamics 
only'' models remain small.  

In other simulations, we optimized protoneutron star convection's chances of 
developing by starting with ``low-test'' initial conditions. These 
conditions were generated by core collapse simulations that incorporated 
``low-test'' neutrino transport, ignoring the important deleptonization 
effects of neutrino--electron scattering (\nes), and the further deleptonization 
accompanying ion--ion screening corrections to the neutral-current coherent 
isoenergetic scattering on nuclei, which dominates the neutrino opacity
during infall. It is well known that less deleptonization occurs when \nes\ 
is neglected, resulting in a larger inner core, a stronger initial shock, 
a larger initial shock radius, and a larger initial entropy gradient
that will subsequently drive a more vigorous protoneutron star convection. Whereas
the initial protoneutron star convection episode in the ``hydrodynamics only'' model
was rather dramatic, with neutrino transport included, the behavior
observed in simulations beginning with realistic initial conditions 
was repeated: The two-dimensional entropy profiles showed very little
protoneutron star convection, and the angle-averaged radial and angular convection
velocities were significantly reduced relative to their values in the
``hydrodynamics only'' model. This occurred even in a final simulation
in which we provided protoneutron star convection with its best chance 
of developing, by starting
with ``low-test'' initial conditions while simultaneously implementing
reduced entropy and lepton equilibration rates to account for neutrino 
diffusion. 

A simple analytical model supports our numerical results, and
demonstrates that neutrino transport reduces the convection growth rates
%and asymptotic velocities by factors of 4--250 for $\rho=10^{11}-10^{12}$
%gcm$^{-3}$, respectively.
and asymptotic velocities by a factor of $\tau_{BV_{s}}/\tau_{s}$, for
entropy-driven convection, and by a factor of $\tau_{BV_{Y_{\ell}}}/\tau_{Y_{\ell}}$, for
lepton-driven convection, where $\tau_{s}$ and $\tau_{Y_{\ell}}$ are the time scales
for equilibrating via neutrino transport a convecting element with the background, in entropy and
leptons, respectively, and $\tau_{BV_{s,Y_{\ell}}}$ are the
e-folding times for convection's growth in the absence of neutrino transport. 
With ``finite-time'' corrections to the equilibration rates, the ratio $\tau_{BV_{s}}/\tau_{s}$
ranges from $\sim$ 3--50 between the neutrinosphere and $10^{12}$ gcm$^{-3}$;                
the ratio $\tau_{BV_{Y_{\ell}}}/\tau_{Y_{\ell}}$ ranges from $\sim$ 250--1000. 

Most prior multidimensional simulations of protoneutron star convection (\eg, Burrows and 
Fryxell 1992, 1993, and Janka and M\"{u}ller 1993b, 1996) have either ignored neutrino 
transport entirely, or have employed radial-ray neutrino transport. While the latter 
is necessitated by the extreme computational challenge imposed by true multidimensional 
transport, it ignores the tendency of neutrino transport to equilibrate upflows and 
downflows with each other in energy and leptons, thus reducing the very buoyancy 
contrast driving these flows. Our {\small 1-D} neutrino transport is complementary 
to radial-ray transport in this regard. Imposing a spherically symmetric neutrino 
distribution is equivalent to maximizing the lateral angular transport of neutrinos 
between ascending high-entropy and descending low-entropy convecting elements, which
would be expected to suppress protoneutron star convection as the rising and falling 
flows equilibrate in entropy and leptons. No doubt differences between our results 
and those of the aforementioned groups derive in part from these two very different 
transport approximations (as well as from our use of multigroup transport and their
use of gray transport, which in semitransparent regions breaks down). 

To assess the validity of our neutrino transport approximation, 
\ie, our use of a precalculated spherically 
symmetric neutrino distribution, we can compare our equilibration times with those 
obtained from detailed neutrino equilibration experiments (Bruenn \etal\ 1995, Bruenn 
\& Dineva 1996, 1997). These are shown in Figure 12. The equilibration time scales 
$\tau_{s}$ and $\tau_{Y_{\ell}}$ correspond to a perturbation of a fluid element in 
entropy and lepton fraction, respectively, and both equilibration time scales 
are computed for a fluid perturbation scale equal to 1 pressure scale height, 
which is $\sim$ 10 km at all densities. For smaller length scales, the equilibration 
time scales would be reduced. As defined in Section 4, the Brunt-V\"{a}is\"{a}l\"{a} time scale, 
$\tau_{BV_{s}}$, is computed for a gradient in $s$ of 1, in 30 km, and no gradient 
in $Y_{\ell}$, and is appropriate for entropy-driven convection. Similarly, 
the Brunt-V\"{a}is\"{a}l\"{a} time scale, $\tau_{BV_{Y_{\ell}}}$, is computed for a gradient 
in $Y_{\ell}$ of 0.1, in 30 km, and no gradient in $s$, and is appropriate for 
lepton-driven convection. Both $\tau_{s}$ and $\tau_{Y_{\ell}}$ exhibit a minimum 
near the neutrinosphere. At higher densities they increase because of slow neutrino 
diffusion, and at low densities they increase because of weak neutrino coupling with 
the matter. 

The equilibration experiments show that a fluid element of one pressure scale height in 
radius will equilibrate in entropy in a minimum time of 3.1 ms, which occurs at 
$\sim 10^{11}$ g cm$^{-3}$ . Smaller modes will equilibrate faster. The lepton 
equilibration time has a minimum of 0.29 ms, which also occurs at $\sim 
10^{11}$ g cm$^{-3}$. On the other hand, our $\dot{\epsilon}$ (with ``finite-time'' 
corrections) gives $\tau_{s}$ values for our 25 \msol\ model that decrease from 0.62 ms at 
the neutrinosphere ($\sim 3 \times 10^{11}$ \gcm), which is $\sim$ 10 times shorter than is 
observed in the equilibration experiments, to 0.03 ms at $10^{12}$ g cm$^{-3}$, which is 
$\sim$ 333 times shorter than is observed in the equilibration experiments. 
Similarly, values for our 15 \msol\ model decrease from 0.65 ms at the 
neutrinosphere to 0.028 ms at $10^{12}$ g cm$^{-3}$. Our $\dot{Y}_{e}$ (with 
``finite-time'' corrections) gives $\tau_{Y_{\ell}}$ values for 
our 25 \msol\ model that decrease from 0.049 ms at the neutrinosphere,
which is $\sim$ 6 times shorter than is observed in the equilibration 
experiments, to 0.018 ms at $10^{12}$ g cm$^{-3}$, which is $\sim$ 38 
times shorter than is observed in the equilibration 
experiments. Similarly, values for our 15 \msol\ model decrease from 
0.054 ms at the neutrinosphere to 0.017 ms at $10^{12}$ g cm$^{-3}$.

Because the density in the initial Ledoux unstable region in our 25 \msol\ 
model ranges from $6.9\times 10^{11}$ g/cm$^{3}$ at the top of the region
to $1.38\times 10^{12}$ g/cm$^{3}$ at the bottom, and in our 15 \msol\ 
model ranges from $5.49\times 10^{11}$ g/cm$^{3}$ to $1.52\times 10^{12}$ 
g/cm$^{3}$, 
at the outset of our simulations we are in a density range in
which, based on our time scale analysis, we may seriously overestimate the 
entropy and lepton equilibration rates, even with ``finite-time'' corrections. 
Whether or not we must revise our conclusions regarding the effect of neutrino 
transport on the growth of convection depend, as we 
discuss below, on the type of protoneutron star convection.

\begin{itemize}
\item {\it Entropy-Driven or Entropy- and Lepton-Driven Convection}:~
As shown in Figure 12, the ratio of the Brunt-V\"{a}is\"{a}l\"{a} time scales 
to the entropy equilibration time scales given by the equilibration 
experiments is always greater than unity, implying that the equilibration time is 
longer than the convection growth time. Therefore, based on this alone, 
there is no doubt that in our approximation we overestimate transport's effect in 
suppressing convection.
%for the case in which the convection is driven by a negative 
%entropy gradient.

\item {\it Lepton-Driven Convection}:~
On the other hand, after the unstable entropy gradients have been eliminated by 
an initial round of entropy-driven convection, or by neutrino diffusion, protoneutron 
star convection tends to continue, driven by a negative lepton fraction gradient 
maintained by ongoing deleptonization at the neutrinosphere. This lepton-driven 
protoneutron star convection is forced by lepton fraction differences between 
convecting fluid elements and the background. For this case, the appropriate 
lepton equilibration time scales are given in Figure 12 by $\tau_{Y_{\ell}}$. 
Figure 12 also shows the appropriate Brunt-V\"{a}is\"{a}l\"{a} time scales, 
$\tau_{BV_{Y_{\ell}}}$. The lepton 
equilibration time scales are significantly smaller than the convection growth times
($\tau_{BV_{Y_{\ell}}}$) for densities between $3 \times 10^{10}$ and $3 \times 10^{12}$ 
\gcm; therefore, in the case of lepton-driven convection,
which is the most important of protoneutron star convection modes, our conclusions 
regarding the tendency of neutrino transport to severely suppress convection are 
realistic.
\end{itemize}

To conclude: There is clearly a need to self-consistently couple multidimensional multigroup
realistic neutrino transport and multidimensional hydrodynamics in optically thick regions 
before we or any other group can draw final conclusions regarding the development of protoneutron 
star convection, or any mode of convection below the neutrinospheres, in core collapse supernovae. 
%In the case of entropy-driven or both entropy- and lepton-driven convection, our use of an imposed 
%spherically symmetric neutrino distribution overestimates neutrino transport's effect in suppressing 
%convection. On the other hand, in the case of lepton-driven convection, which is the most important 
%case, our use of an imposed spherically symmetric distribution is a better approximation. 
%
%Our use of a spherically symmetric neutrino 
%distribution, complementary to ray-by-ray schemes, maximizes the equilibrating neutrino transport 
%one should expect between upflows and downflows in a fully multidimensional transport simulation, 
%and consequently, suppresses convection more than should be expected. 
Nonetheless, our compendium 
of results, supported by our analytical model and time scale analyses, is compelling, pointing to a 
very real effect: Neutrino 
transport will suppress protoneutron star convection, the extent of which lies somewhere between 
what we have computed and what other groups have computed. The question is: Where? 
\newpage

\section{Acknowledgements}

AM, ACC, MWG, and MRS were supported at the Oak Ridge National 
Laboratory, which is managed by Lockheed Martin Energy Research 
Corporation under DOE contract DE-AC05-96OR22464. AM, MWG, and MRS were 
supported at the University of Tennessee under DOE contract 
DE-FG05-93ER40770. ACC and SU were supported at Vanderbilt 
University under DOE contract DE-FG02-96ER40975. SWB was 
supported at Florida Atlantic University under NSF grant 
AST--9618423 and NASA grant NRA-96-04-GSFC-073, 
and JMB was supported at North Carolina State 
University under NASA grant NAG5-2844. 

The simulations presented in this paper were carried out on
the Cray C90 at the National Energy Research Supercomputer
Center, the Cray Y/MP at the North Carolina Supercomputer 
Center, and the Cray Y/MP and Silicon Graphics Power
Challenge at the Florida Supercomputer Center. 

We would like to thank 
%Adam Burrows
%, 
%Chris Fryer
%, 
%Wolfgang Hillebrandt
%, 
Thomas Janka
%, 
%Ewald M\"{u}ller
%, 
%Doug Swesty
%, 
%and Friedel Thielemann 
for stimulating discussions, and the referee, Marc Herant, for many useful 
criticisms and suggestions that considerably expanded and strengthened the 
content of this work. 
%We would also like to thank Michael Smith for carefully
%reading the manuscript. 
\newpage

\section{\bf References}

%\begin{enumerate}
\noindent
%\item 
Arnett, W. D. 1986, in IAU Symposium 125, The Origin and Evolution of
      Neutron\linebreak\indent Stars, ed. D. J. Helfand and J. H. Huang (Dordrecht: Reidel) 

\noindent
%\item 
Arnett, W. D. 1987, ApJ, 319, 136 

\noindent
%\item 
Bazan, G. \& Arnett, W. D. 1994, ApJ, 433, L41 

\noindent
%\item 
Bethe, H. A. 1990, Rev. Mod. Phys., 62, 801 

\noindent
%\item 
Bethe, H. A. 1993, ApJ, 412, 192 

\noindent
%\item 
Bethe, H. A., Brown, G. E., \& Cooperstein, J. 1987, ApJ, 322, 201 

\noindent
%\item 
Bethe, H. \& Wilson, J. R. 1985, ApJ, 295, 14

\noindent
%\item 
Bruenn, S. W. 1993, in Nuclear Physics in the Universe, eds. M. W. 
      Guidry and\linebreak\indent M. R. Strayer (IOP Publishing, Bristol), p. 31

\noindent
%\item 
Bruenn, S. W., Buchler, J. R., \& Livio, M. 1979, ApJ, 234, L183 

\noindent
%\item 
Bruenn, S. W. \& Dineva, T. 1996, ApJ, 458, L71

\noindent
%\item 
Bruenn, S. W. \& Dineva, T. 1997, in preparation 

\noindent
%\item 
Bruenn, S. W. \& Mezzacappa, A. 1994, ApJ, 433, L45

\noindent
%\item 
Bruenn, S. W., Mezzacappa, A., \& Dineva, T. 1995, Phys. Rep., 256, 69

\noindent
%\item 
Burrows, A. 1987, ApJ, 318, L63

\noindent
%\item 
Burrows, A. \& Fryxell, B. A. 1992, Science, 258, 430

\noindent
%\item 
Burrows, A. \& Fryxell, B. A. 1993, ApJ, 418, L33

\noindent
%\item 
Burrows, A., Hayes, J., \& Fryxell, B. A. 1995, ApJ, 450, 830

\noindent
%\item 
Burrows, A. \& Lattimer, J. M. 1988, Phys. Rep., 163, 5

\noindent
%\item 
Colgate, S. A., Herant, M., \& Benz, W. 1993, Phys. Rep., 227, 157 

\noindent
%\item 
Colgate, S. A., \& Petschek, A. G. 1980, ApJ, 236, L115 

\noindent
%\item 
Cooperstein, J. 1993, in Nuclear Physics in the Universe, eds. M. W. 
      Guidry and\linebreak\indent M. R. Strayer (IOP Publishing, Bristol), p. 99

\noindent
%\item 
Epstein, R. I. 1979, MNRAS, 188, 305 

\noindent
%\item 
Herant, M., Benz, W., \& Colgate, S. A. 1992, ApJ, 395, 642

\noindent
%\item 
Herant, M., Benz, W., Hix, W. R., Fryer, C. L., \& Colgate, S. A. 1994,
      ApJ, 435, 339

\noindent
%\item 
Janka, H.-Th. \& M\"{u}ller, E. 1993a, in Proc. of the IAU Coll. 
      145 (Xian China, May\linebreak\indent 24--29, 1993), eds. R. McCray and Wang Zhenru, 
      (Cambridge Univ. Press, Cambridge);\linebreak\indent MPA-Preprint 748 

\noindent
%\item 
Janka, H.-Th. \& M\"{u}ller, E. 1993b, in Frontiers of Neutrino 
      Astrophysics, eds. Y. Suzuki\linebreak\indent and K. Nakamura, (Universal Academy Press, 
      Tokyo). p. 203 

\noindent
%\item 
Janka, H.-Th. \& M\"{u}ller, E. 1995, ApJ, 448, L109 

\noindent
%\item 
Janka, H.-Th., \& M\"{u}ller, E. 1996, A\&A 306, 167

\noindent
%\item 
Keil, W., Janka, H.-Th, \& M\"{u}ller, E. 1996, ApJ, 473, L111

\noindent
%\item 
Lattimer, J. M., \& Mazurek, T. J. 1981, ApJ, 246, 995 

\noindent
%\item 
Lattimer, J. M. \& Swesty , F. D. 1991, Nucl. Phys. A, 535, 331

\noindent
%\item 
Livio, M., Buchler, J. R., \& Colgate, S. A. 1980, ApJ, 238, L139 

\noindent
%\item 
Mayle, R. W. 1985, Ph.D. Thesis, Univ. California, Berkeley (UCRL preprint 
      no. 53713)

\noindent
%item
Mezzacappa, A. \& Bruenn, S. W. 1993, ApJ, 405, 669

\noindent
%\item 
Mezzacappa, A., Calder, A. C., Bruenn, S. W., Blondin, J. M.,
      Guidry, M. W.,\linebreak\indent Strayer, M. R., \& Umar, A. S. 1997, ApJ, in press 

\noindent
%\item 
Miller, D. S., Wilson, J. R., \& Mayle, R. W. 1993, ApJ, 415, 278

\noindent
%\item 
M\"{u}ller, E. 1993, in Proc. of the 7th Workshop on Nuclear Astrophysics 
      (Ringberg Castle,\linebreak\indent March 22--27, 1993), eds. W. Hillebrandt and E. 
      M\"{u}ller, Report MPA/P7,\linebreak\indent Max-Plank-Institut f\"{u}r Astrophysik, 
      Garching, p. 27 

\noindent
%\item 
M\"{u}ller, E. \& Janka, H.-Th. 1994, in Reviews in Modern Astronomy 7 
     (Proc. of the Int.\linebreak\indent Conf. of the AG (Bochum, Germany, 1993)), eds. G. Klure 
     (Astronomische Gesellschaft,\linebreak\indent Hamburg), p. 103 

\noindent
%\item 
Shimizu, T., Yamada, S., \& Sato, K. 1993, Publ. Astron. Soc. Japan, 45, L53

\noindent
%\item 
Shimizu, T., Yamada, S., \& Sato, K. 1994, ApJ, 432, L119

\noindent
%\item 
Smarr, L., Wilson, J. R., Barton, R. T., \& Bowers, R. L. 1981 ApJ, 
      246, 515

\noindent
%\item 
Weaver, T. A. \& Woosley, S. E. 1997, ApJ, in preparation 

\noindent
%\item 
Wilson, J. R. 1985, in Numerical Astrophysics, eds. J. M. Centrella \etal\
      (Boston: Jones\linebreak\indent \& Bartlett), 422

\noindent
%\item 
Wilson, J. R. \& Mayle, R. W. 1988, Phys. Rep., 163, 63

\noindent
%\item 
Wilson, J. R. \& Mayle, R. W. 1993, Phys. Rep., 227, 97

\noindent
%\item 
Wilson, J. R., Mayle, R. W., Woosley, S. E., \& Weaver, T. 1986, 
      Ann. N.Y. Acad.\linebreak\indent Sci., 470, 367

\noindent
%\item 
Woosley, S. E. \& Weaver, T. A. 1995, ApJS, 101, 181 

%\end{enumerate}
\newpage

\begin{table}
%\centering
%\begin{tabular}{||c|c|c|l||}        \hline
\begin{tabular}{cccl}        \hline
%\begin{tabular}{cccc}        \hline
{\bf Simulation} & {\bf Initial Time}  & {\bf Postbounce Time}
 & {\bf Description}  \\
  & (ms) & (ms) & \\
\hline  \hline
%\multicolumn{4}{||c||}{{\bf 25 ${\rm M}_{\odot}$ Models }} \\ \hline
\multicolumn{4}{c}{{\bf 25 ${\rm M}_{\odot}$ Models }} \\ \hline
%\multicolumn{3}{|l|}{ 25}  \\ \hline
A & 297.9 & 8.3 & 1-d hydro only \\ \hline
B & 297.9 & 8.3 & 1-d hydro + $\nu$ transport \\ \hline
C & 297.9 & 8.3 & 2-d hydro only \\ \hline
D & 297.9 & 8.3 & 2-d hydro + $\nu$ transport \\ \hline
E & 297.9 & 8.3 & 2-d ``Finite Time" hydro + $\nu$ \\ \hline
F & 299.4 & 7.23 & 2-d ``Low Test" hydro only  \\ \hline
G & 299.4 & 7.23 & 2-d ``Low Test" hydro + $\nu$ \\ \hline
H & 299.4 & 7.23 & 2-d ``F. T. L. T." hydro + $\nu$ \\ \hline \hline
%\multicolumn{4}{||c||}{{\bf 15 ${\rm M}_{\odot}$ Models }} \\ \hline
\multicolumn{4}{c}{{\bf 15 ${\rm M}_{\odot}$ Models }} \\ \hline
I & 211.1 & 11.9 & 1-d hydro only \\ \hline
J & 211.1 & 11.9 & 1-d hydro + $\nu$ transport \\ \hline
K & 211.1 & 11.9 & 2-d hydro only \\ \hline
L & 211.1 & 11.9 & 2-d hydro + $\nu$ transport \\ \hline
\end{tabular}
\caption{Simulation ``start times'' and descriptions.}
\end{table}
\newpage

\begin{table}
\centering
%\begin{tabular}{||c|c|c|c|c|c|c||}      
\begin{tabular}{ccccccc}      
  \hline
{\bf Simulation} 
%& {\bf Seeding Radii} (Km)  & {\bf Entropy} ($s = S/kN_B$) 
%& {\bf Electron Fraction} 
%} \\ \hline 
%& \multicolumn{2}{|c|} {\hspace{9pt} {\bf Seeding Radii}\hspace{9pt} }  
& \multicolumn{2}{c} {\hspace{9pt} {\bf Seeding Radii}\hspace{9pt} }  
%& \multicolumn{2}{|c|} {\hspace{25.5pt} {\bf Entropy}\hspace{25.5pt} }  
& \multicolumn{2}{c} {\hspace{25.5pt} {\bf Entropy}\hspace{25.5pt} }  
%& \multicolumn{2}{|c||} {  {\bf Electron Fraction} }  
& \multicolumn{2}{c} {  {\bf Electron Fraction} }  
\\ 
%& \multicolumn{2}{|c|} {   (Km) }
& \multicolumn{2}{c} {   (km) }
%& \multicolumn{2}{|c|} {  (s = S/kN$_{\rm B}$) }  
& \multicolumn{2}{c} {  (s = S/kN$_{\rm B}$) }  
%& \multicolumn{2}{|c||} {   }  
& \multicolumn{2}{c} {   }  
\\ \hline 
&\hspace{13.5pt}$r_{\rm min}$\hspace{13.5pt} 
& $r_{\rm max}$ 
&\hspace{13.0pt}s$_{\rm  min}$\hspace{13.0pt} 
%& $s_{\rm min}$ 
& $s_{\rm  max}$ 
&\hspace{9pt}$\left(Y_{\rm e}\right)_{\rm min }$\hspace{9pt} 
& $\left(Y_{\rm e}\right)_{\rm max }$ \\ \hline \hline
%\multicolumn{7}{||c||}{{\bf 25 ${\rm M}_{\odot}$ Models }} \\ \hline 
\multicolumn{7}{c}{{\bf 25 ${\rm M}_{\odot}$ Models }} \\ \hline 
A & 33.8 & 52.0 & 5.08  & 6.24 & 0.118 & 0.221 \\ \hline
B & 33.8 & 52.0 & 5.08  & 6.24 & 0.118 & 0.221 \\ \hline
C & 33.8 & 52.0 & 5.08  & 6.24 & 0.118 & 0.221 \\ \hline
D & 33.8 & 52.0 & 5.08  & 6.24 & 0.118 & 0.221 \\ \hline
E & 33.8 & 52.0 & 5.08  & 6.24 & 0.118 & 0.221 \\ \hline
F & 55.4 & 122.2 & 3.33 & 7.74 & 0.213 & 0.463 \\ \hline
G & 55.4 & 122.2 & 3.33 & 7.74 & 0.213 & 0.463 \\ \hline
H & 55.4 & 122.2 & 3.33 & 7.74 & 0.213 & 0.463 \\ \hline \hline
%\multicolumn{7}{||c||}{{\bf 15 ${\rm M}_{\odot}$ Models }} \\ \hline 
\multicolumn{7}{c}{{\bf 15 ${\rm M}_{\odot}$ Models }} \\ \hline 
I & 31.0 & 54.2 & 5.25  & 6.19 & 0.116 & 0.234 \\ \hline
J & 31.0 & 54.2 & 5.25  & 6.19 & 0.116 & 0.234 \\ \hline
K & 31.0 & 54.2 & 5.25  & 6.19 & 0.116 & 0.234 \\ \hline
L & 31.0 & 54.2 & 5.25  & 6.19 & 0.116 & 0.234 \\ \hline
\end{tabular}
\caption{Radial ranges, and entropy and electron fraction maxima and minima, for the
initial Ledoux unstable regions.}
\end{table}

\end{document}

\newcounter{numbercount}
\newcounter{lettercount}

\setcounter{numbercount}{1}
\setcounter{lettercount}{1}
\setcounter{page}{46}

% Figure page 1.

\begin{figure}
\centering
\epsfig{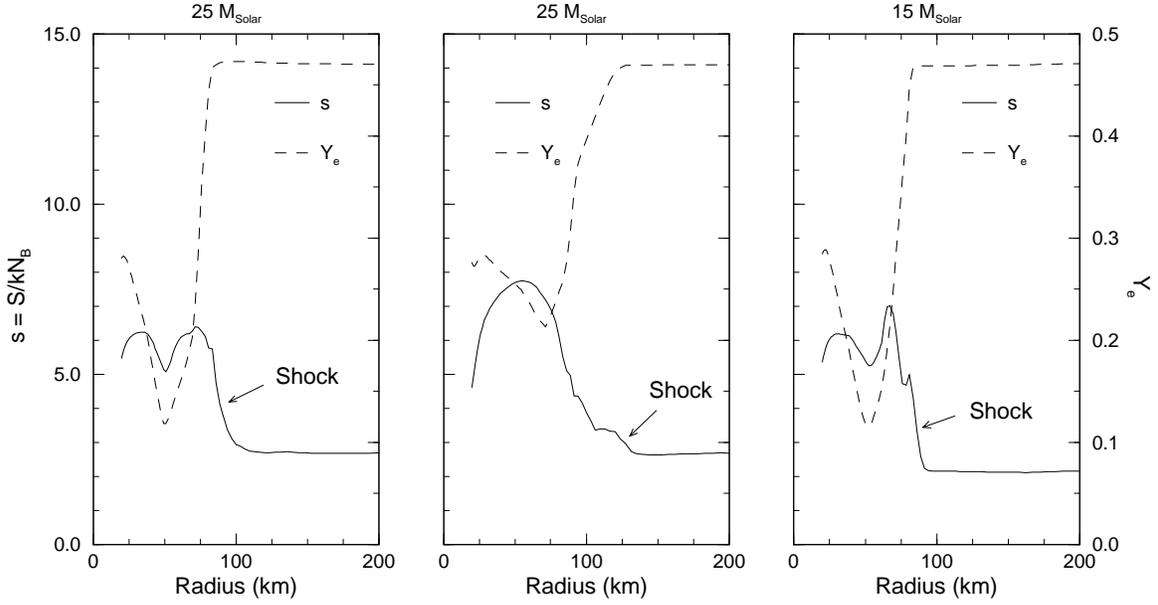}
\caption[Initial entropy and electron fraction profiles]
{\protect\small \protect\baselineskip 10pt \label{pfig1}
Initial entropy and electron fraction profiles of the
25 ${\rm M}_{\odot}$, 25 ${\rm M}_{\odot}$ ``low--test",
and 15 ${\rm M}_{\odot}$ simulations.}
\end{figure}
 
\addtocounter{numbercount}{1}
\setcounter{lettercount}{1}
\renewcommand{\thefigure}{\arabic{numbercount}\alph{lettercount}}

\begin{figure}
\centering
\epsfig{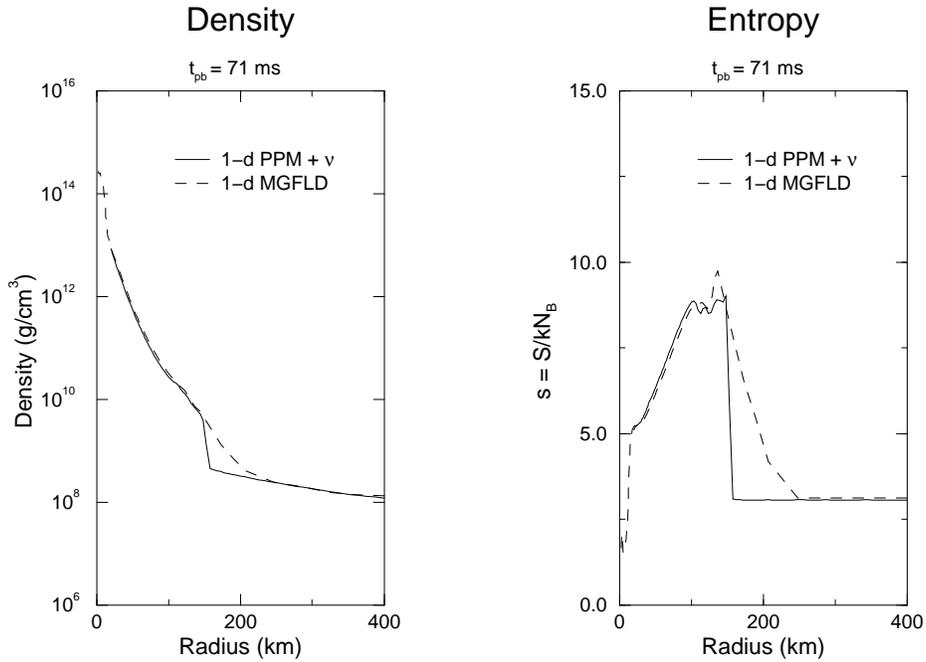}
\caption[Initial entropy and electron fraction profiles]
{\protect\small \protect\baselineskip 10pt \label{pfig2a}
Plot of density and entropy versus radius,
comparing a simulation with one-dimensional
{\small PPM} hydrodynamics and {\small MGFLD} neutrino transport
(simulation B) to a
simulation with one-dimensional Lagrangian hydrodynamics and
{\small MGFLD} neutrino transport. The comparison is made 62
ms into the run.}
\end{figure}

\clearpage

% Figure page 2.

\addtocounter{lettercount}{1}

\begin{figure}
\centering
\epsfig{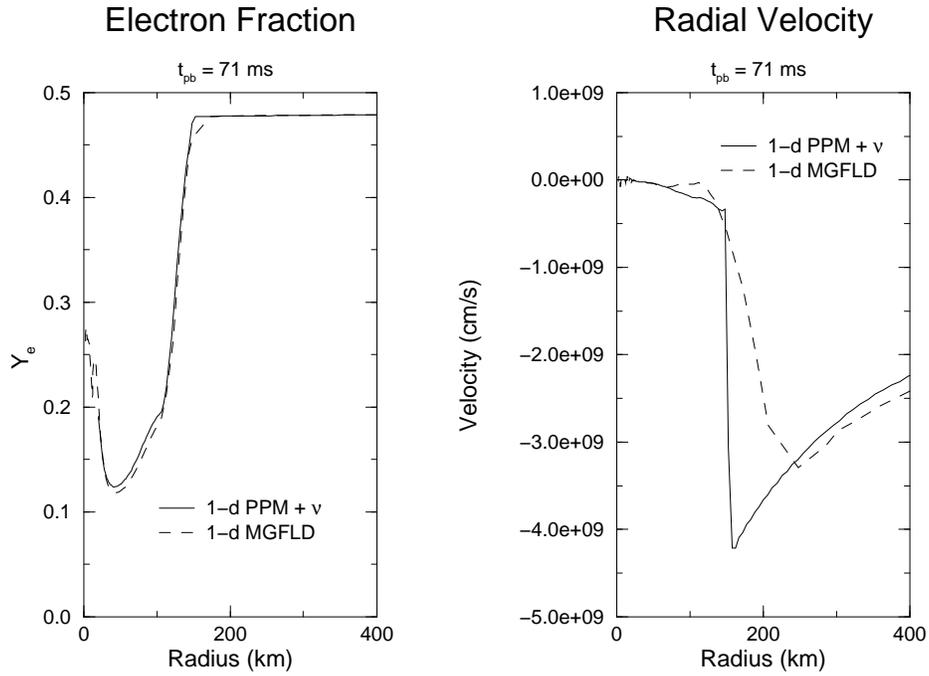}
\caption[Electron fraction and radial velocity vs. radius]
{\protect\small \protect\baselineskip 10pt \label{pfig2b}
Plot of electron fraction and radial velocity vs. radius
for the same comparison shown in Figure~\ref{pfig2a}.}
\end{figure}

\addtocounter{lettercount}{1}

\begin{figure}
\centering
\epsfig{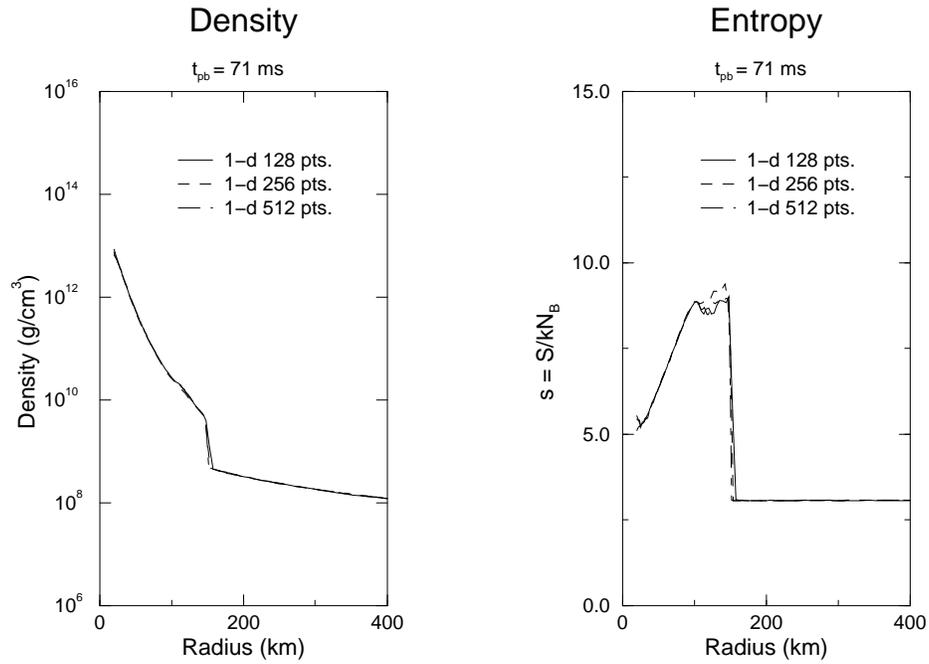}
\caption[Density and  entropy vs. radius]
{\protect\small \protect\baselineskip 10pt \label{pfig2c}
Plots of the density and entropy profiles at $t_{\rm pb}=71$ ms,
for three different radial grid resolutions in our one-dimensional
{\small PPM} simulations.}
\end{figure}

\clearpage

% Figure page 3.

\addtocounter{lettercount}{1}

\begin{figure}
\centering
\epsfig{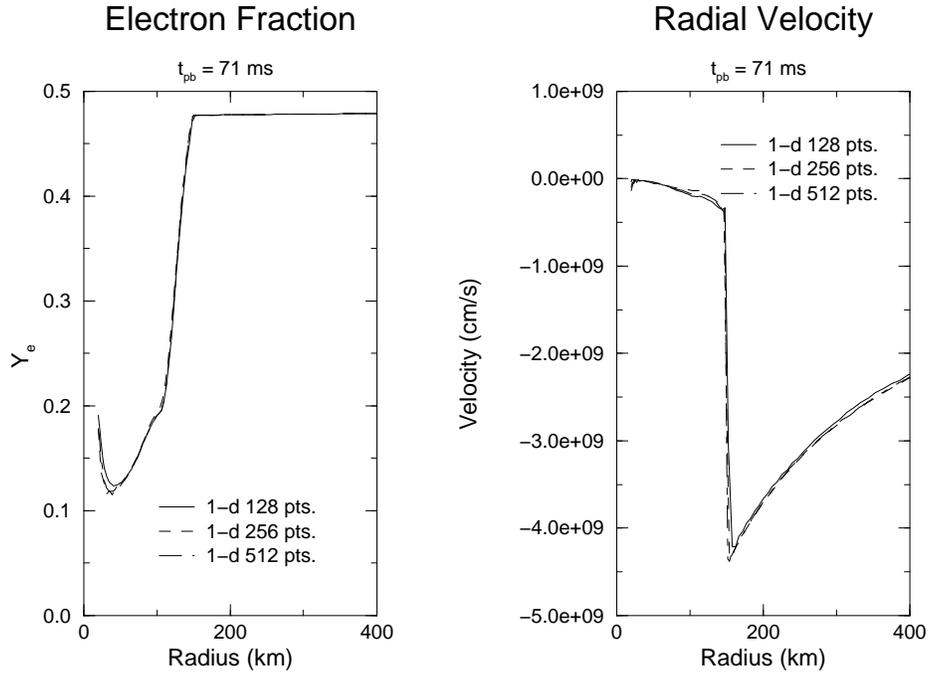}
\caption[Electron fraction and radial velocity vs. radius]
{\protect\small \protect\baselineskip 10pt \label{pfig2d}
Plots of the electron fraction and radial velocity profiles at
$t_{\rm pb}=71$ ms, for three different radial grid resolutions in
our one-dimensional {\small PPM} simulations.}
\end{figure}

\addtocounter{numbercount}{1}
\setcounter{lettercount}{2}

\begin{figure}
\centering
\epsfig{file=pfigs/pfig3b.epsi,width=6.0in,angle=0}
\caption[Angle-averaged radial and angular convection velocities]
{\protect\small \protect\baselineskip 10pt \label{pfig3b}
Plots of the angle-averaged radial and angular convection velocities
corresponding to the three entropy snapshots in Figure~\ref{pfig3a}.}
\end{figure}

\clearpage

% Figure page 4.

\addtocounter{lettercount}{1}

\begin{figure}
\centering
\epsfig{file=pfigs/pfig3c.epsi,width=6.0in,angle=0}
\caption[Initial entropy and electron fraction profiles]
{\protect\small \protect\baselineskip 10pt \label{pfig3c}
Plots showing the angle-averaged entropy and
electron fraction corresponding to the three
entropy snapshots in Figure~\ref{pfig3a}.}
\end{figure}

\addtocounter{numbercount}{1}
\setcounter{lettercount}{2}

\begin{figure}
\centering
\epsfig{file=pfigs/pfig4b.epsi,width=6.0in,angle=0}
\caption[Angle averaged radial and angular convection velocities]
{\protect\small \protect\baselineskip 10pt \label{pfig4b}
Plots of the angle-averaged radial and angular convection velocities
corresponding to the three entropy snapshots in Figure~\ref{pfig4a}.}
\end{figure}

\clearpage

% Figure page 5.

\addtocounter{lettercount}{1}

\begin{figure}
\centering
\epsfig{file=pfigs/pfig4c.epsi,width=6.0in,angle=0}
\caption[Angle averaged entropy and electron fraction]
{\protect\small \protect\baselineskip 10pt \label{pfig4c}
Plots showing the angle-averaged entropy and electron 
fraction corresponding to the three entropy snapshots 
in Figure~\ref{pfig4a}.}
\end{figure}

\addtocounter{lettercount}{1}

\begin{figure}
\centering
\epsfig{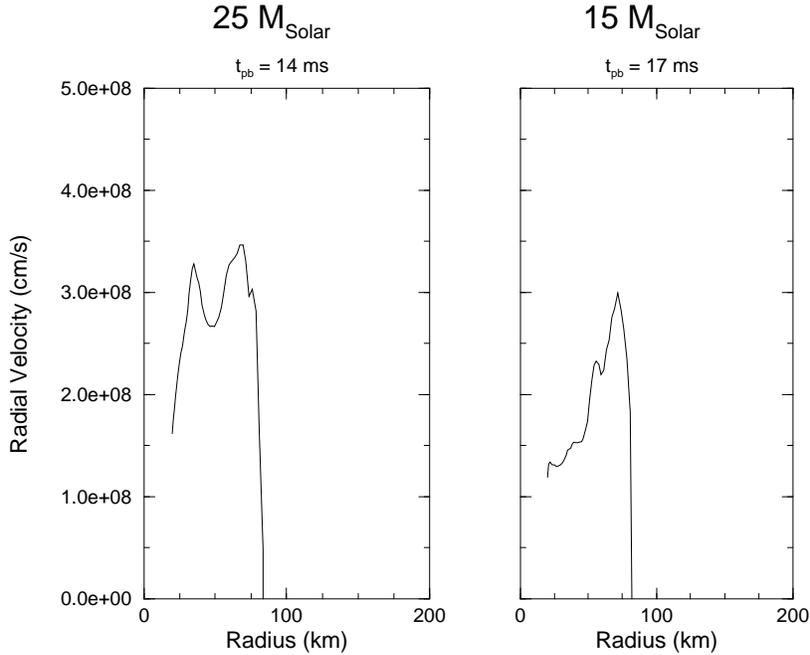}
\caption[Angle-averaged radial velocities]
{\protect\small \protect\baselineskip 10pt \label{pfig4d}
Plot of the angle-averaged radial velocities in the innermost 
region of the two-dimensional grid for our ``fiducial" 25 
${\rm M}_{\odot}$ and 15 ${\rm M}_{\odot}$ simulations (D 
and L).}
\end{figure}

\clearpage

% Figure page 6.

\addtocounter{numbercount}{1}
\setcounter{lettercount}{2}

\begin{figure}
\centering
\epsfig{file=pfigs/pfig5b.epsi,width=6.0in,angle=0}
\caption[Angle averaged radial and angular convection velocities]
{\protect\small \protect\baselineskip 10pt \label{pfig5b}
Plots of the angle-averaged radial and angular convection 
velocities corresponding to the three entropy snapshots in
Figure~\ref{pfig5a}.}
\end{figure}

\addtocounter{lettercount}{1}

\begin{figure}
\centering
\epsfig{file=pfigs/pfig5c.epsi,width=6.0in,angle=0}
\caption[Angle averaged entropy and electron fraction]
{\protect\small \protect\baselineskip 10pt \label{pfig5c}
Plots showing the angle-averaged entropy and
electron fraction corresponding to the three
entropy snapshots in Figure~\ref{pfig5a}.}
\end{figure}

\clearpage

% Figure page 7.

\addtocounter{numbercount}{1}
\setcounter{lettercount}{2}

\begin{figure}[p]
\centering
\epsfig{file=pfigs/pfig6b.epsi,width=6.0in,angle=0}
\caption[Angle averaged radial and angular convection velocities]
{\protect\small \protect\baselineskip 10pt \label{pfig6b}
Plots of the angle-averaged radial and angular convection 
velocities corresponding to the three entropy snapshots in
Figure~\ref{pfig6a}.}
\end{figure}

\addtocounter{lettercount}{1}

\begin{figure}[p]
\centering
\epsfig{file=pfigs/pfig6c.epsi,width=6.0in,angle=0}
\caption[Angle averaged entropy and electron fraction]
{\protect\small \protect\baselineskip 10pt \label{pfig6c}
Plots showing the angle-averaged entropy and electron 
fraction corresponding to the three entropy snapshots in 
Figure~\ref{pfig6a}.}
\end{figure}

\clearpage

% Figure page 8.

\addtocounter{numbercount}{1}
\setcounter{lettercount}{2}

\begin{figure}[p]
\centering
\epsfig{file=pfigs/pfig7b.epsi,width=6.0in,angle=0}
\caption[Angle averaged radial and angular convection velocities]
{\protect\small \protect\baselineskip 10pt \label{pfig7b}
Plots of the angle-averaged radial and angular convection 
velocities corresponding to the three entropy snapshots in
Figure~\ref{pfig7a}.}
\end{figure}

\addtocounter{lettercount}{1}

\begin{figure}[p]
\centering
\epsfig{file=pfigs/pfig7c.epsi,width=6.0in,angle=0}
\caption[Angle averaged entropy and electron fraction]
{\protect\small \protect\baselineskip 10pt \label{pfig7c}
Plots showing the angle-averaged entropy and
electron fraction corresponding to the three
entropy snapshots in Figure~\ref{pfig7a}.}
\end{figure}

\clearpage

% Figure page 9.

\addtocounter{numbercount}{1}
\setcounter{lettercount}{2}

\begin{figure}[p]
\centering
\epsfig{file=pfigs/pfig8b.epsi,width=6.0in,angle=0}
\caption[Angle averaged radial and angular convection velocities]
{\protect\small \protect\baselineskip 10pt \label{pfig8b}
Plots of the angle-averaged radial and angular convection 
velocities corresponding to the three entropy snapshots in
Figure~\ref{pfig8a}.}
\end{figure}

\addtocounter{lettercount}{1}

\begin{figure}[p]
\centering
\epsfig{file=pfigs/pfig8c.epsi,width=6.0in,angle=0}
\caption[Angle averaged entropy and electron fraction]
{\protect\small \protect\baselineskip 10pt \label{pfig8c}
Plots showing the angle-averaged entropy and
electron fraction corresponding to the three
entropy snapshots in Figure~\ref{pfig8a}.}
\end{figure}

\clearpage

% Figure page 10.

% Skip figure 9, which is a color figure.
\addtocounter{numbercount}{2}
\setcounter{lettercount}{2}

\begin{figure}[p]
\centering
\epsfig{file=pfigs/pfig10b.epsi,width=6.0in,angle=0}
\caption[Angle averaged radial and angular convection velocities]
{\protect\small \protect\baselineskip 10pt \label{pfig10b}
Plots of the angle-averaged radial and angular convection 
velocities corresponding to the three entropy snapshots in
Figure~\ref{pfig10a}.}
\end{figure}

\addtocounter{lettercount}{1}

\begin{figure}[p]
\centering
\epsfig{file=pfigs/pfig10c.epsi,width=6.0in,angle=0}
\caption[Angle averaged entropy and electron fraction]
{\protect\small \protect\baselineskip 10pt \label{pfig10c}
Plots showing the angle-averaged entropy and
electron fraction corresponding to the three
entropy snapshots in Figure~\ref{pfig10a}.}
\end{figure}

\clearpage

% Figure page 11.

\addtocounter{numbercount}{1}
\setcounter{lettercount}{2}

\begin{figure}
\centering
\epsfig{file=pfigs/pfig11b.epsi,width=6.0in,angle=0}
\caption[Angle averaged radial and angular convection velocities]
{\protect\small \protect\baselineskip 10pt \label{pfig11b}
Plots of the angle-averaged radial and angular convection 
velocities corresponding to the three entropy snapshots in
Figure~\ref{pfig11a}.}
\end{figure}

\addtocounter{lettercount}{1}

\begin{figure}
\centering
\epsfig{file=pfigs/pfig11c.epsi,width=6.0in,angle=0}
\caption[Angle averaged entropy and electron fraction]
{\protect\small \protect\baselineskip 10pt \label{pfig11c}
Plots showing the angle-averaged entropy and
electron fraction corresponding to the three
entropy snapshots in Figure~\ref{pfig11a}.}
\end{figure}

\clearpage

% Figure page 12.

\renewcommand{\thefigure}{\arabic{numbercount}}
\addtocounter{numbercount}{1}

\begin{figure}[t]
\centering
\epsfig{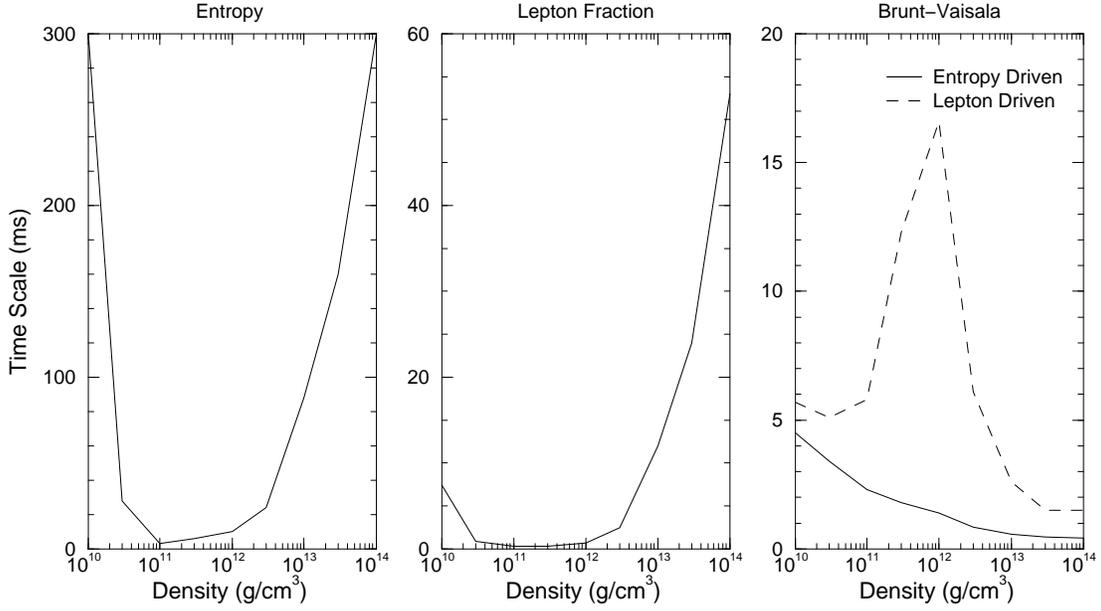}
\caption[Convection and Equilibration Time Scales]
{\protect\small \protect\baselineskip 10pt \label{pfig12}
Plots showing the entropy and lepton-fraction equilibration time
scales, and the convection (Brunt--V\"{a}is\"{a}l\"{a}) time scales,
as a function of density.}
\end{figure}

\renewcommand{\thefigure}{\arabic{numbercount}\alph{lettercount}} 
\setcounter{numbercount}{3}
\setcounter{lettercount}{1}

\begin{figure}
\centering
\hbox{
\epsfig{file=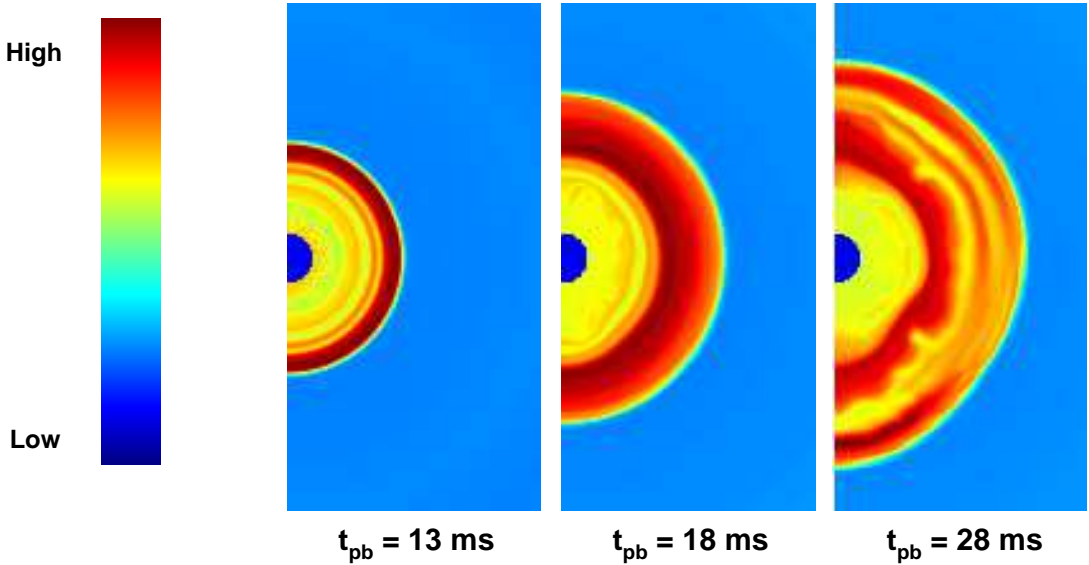,width=5.7in,angle=0}
}
\caption[Two-dimensional entropy (simulation C)]
{\protect\small \protect\baselineskip 10pt \label{pfig3a}
(simulation C).
The two-dimensional plots showing the entropy evolution of the
25 ${\rm M}_{\odot}$  model in a simulation without neutrino 
transport (simulation C).}
\end{figure}

\clearpage

% Figure page 13.

\addtocounter{numbercount}{1}

\begin{figure}
\centering
\hbox{
\epsfig{file=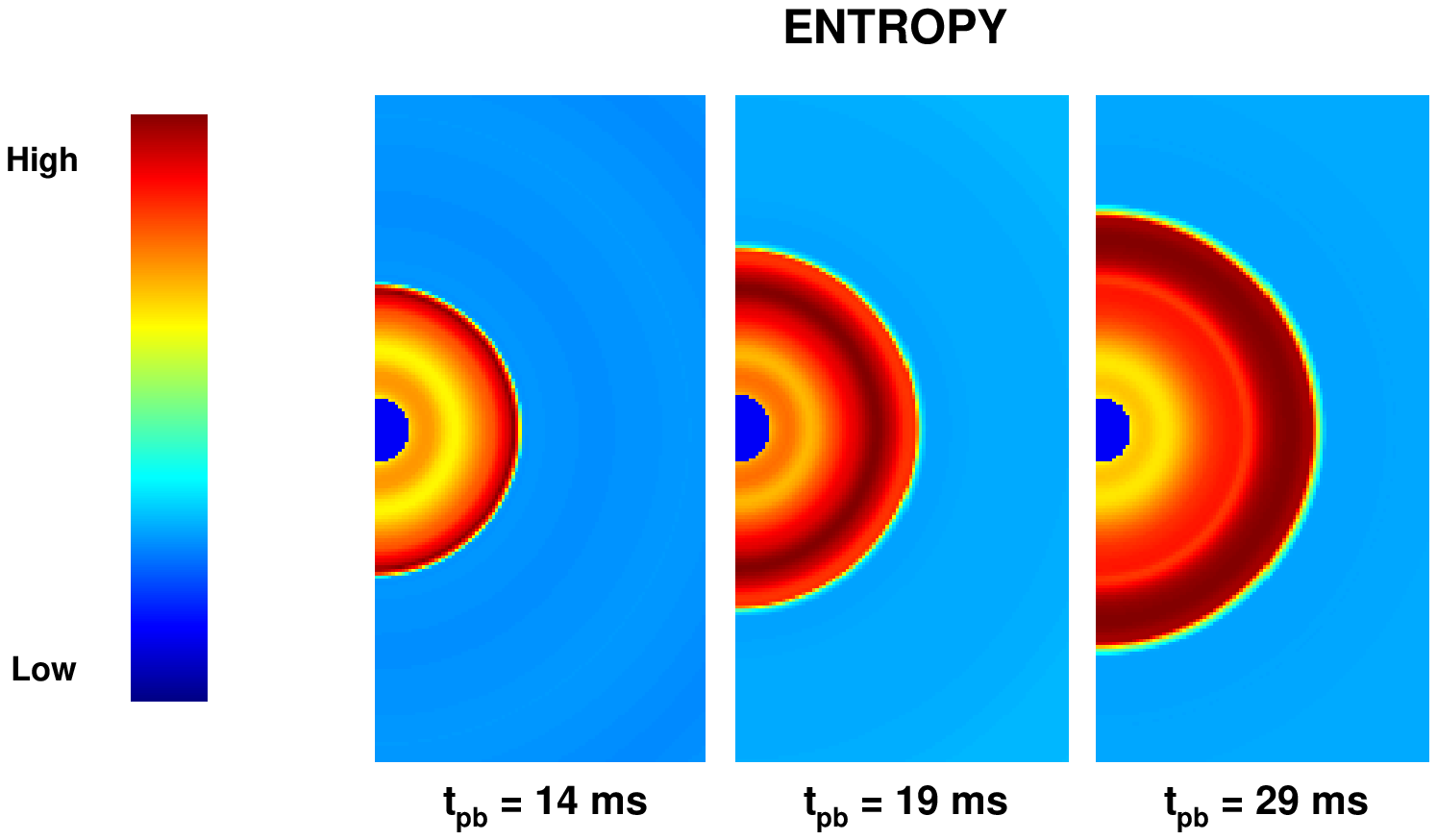,width=5.7in,angle=0}
}
\caption[Two-dimensional entropy (simulation D)]
{\protect\small \protect\baselineskip 10pt \label{pfig4a}
Two-dimensional plots showing the entropy evolution of the
25 ${\rm M}_{\odot}$  model in a simulation with neutrino
transport (simulation D).}
\end{figure}

\addtocounter{numbercount}{1}

\begin{figure}[p]
\centering
\epsfig{file=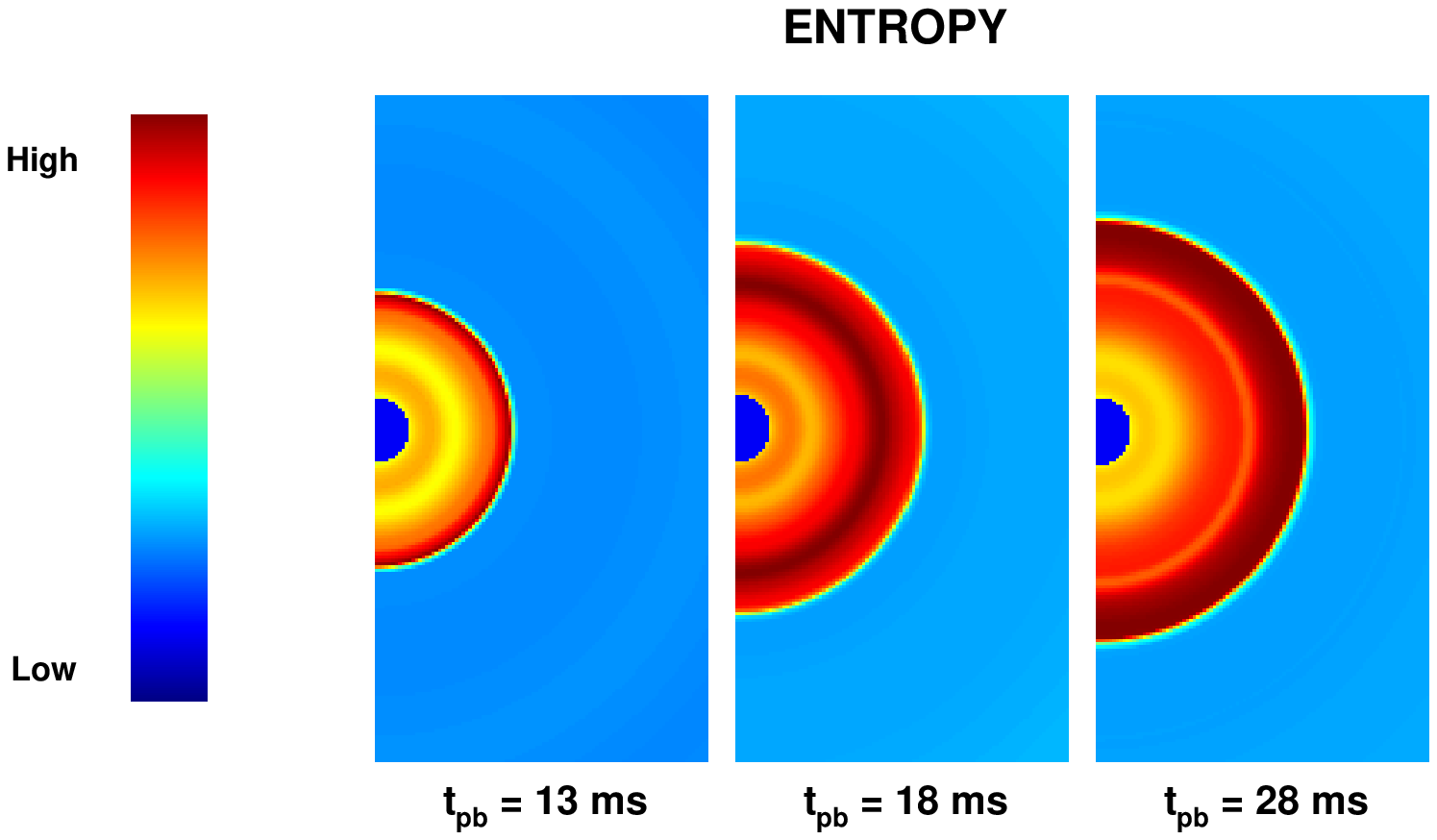,width=5.7in,angle=0}
\caption[Two-dimensional entropy (simulation E)]
{\protect\small \protect\baselineskip 10pt \label{pfig5a}
Two-dimensional plots showing the entropy evolution of the
25 ${\rm M}_{\odot}$  model in a simulation with neutrino
transport that includes ``finite-time" effects (simulation E).}
\end{figure}

\clearpage

% Figure page 14.

\addtocounter{numbercount}{1}

\begin{figure}[p]
\centering
\epsfig{file=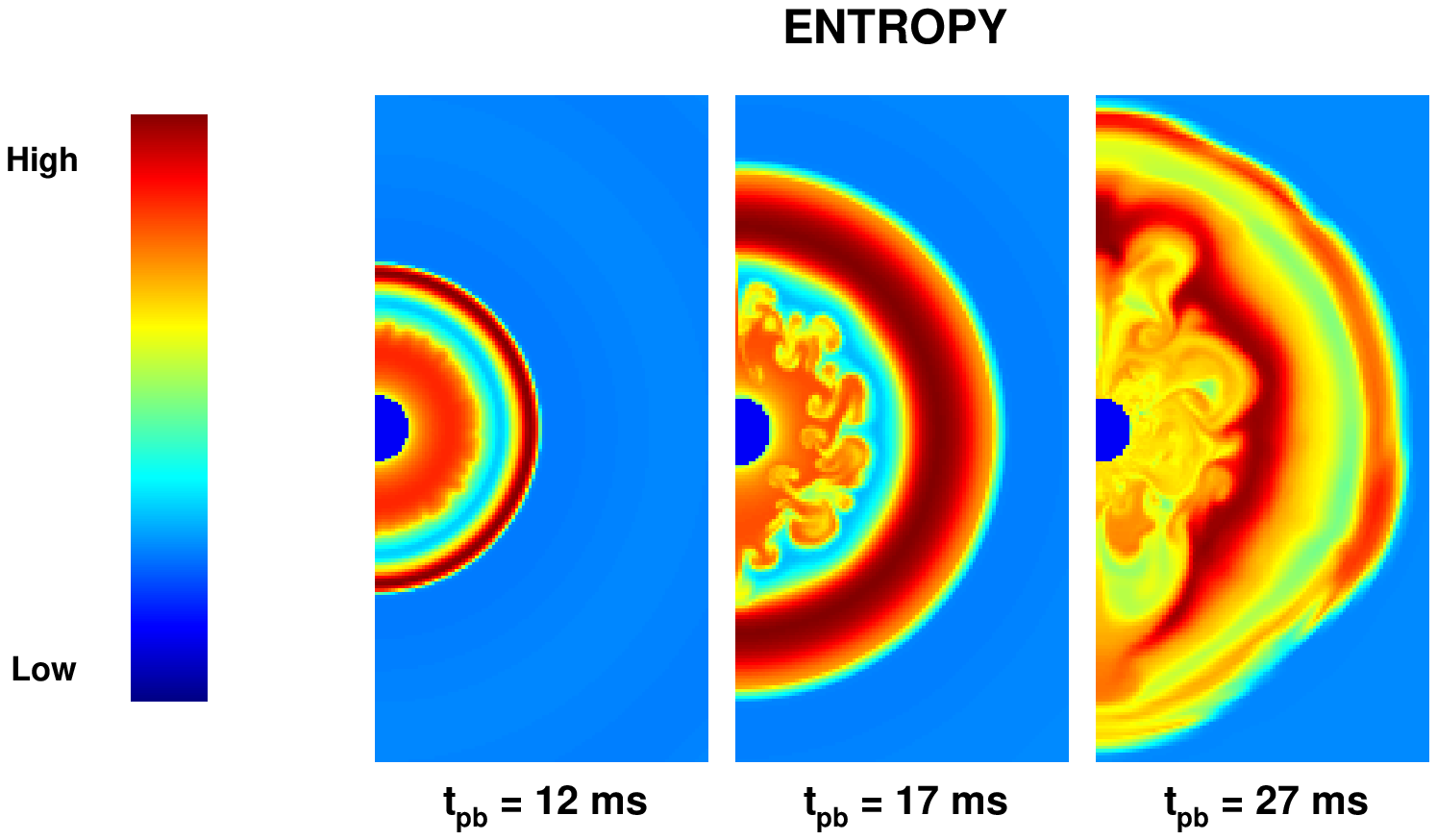,width=5.7in,angle=0}
\caption[Two-dimensional entropy (simulation F)]
{\protect\small \protect\baselineskip 10pt \label{pfig6a}
Two-dimensional plots showing the entropy evolution of the
25 ${\rm M}_{\odot}$  model in a ``low--test" simulation without
neutrino transport (simulation F).}
\end{figure}

\addtocounter{numbercount}{1}

\begin{figure}[tp]
\centering
\epsfig{file=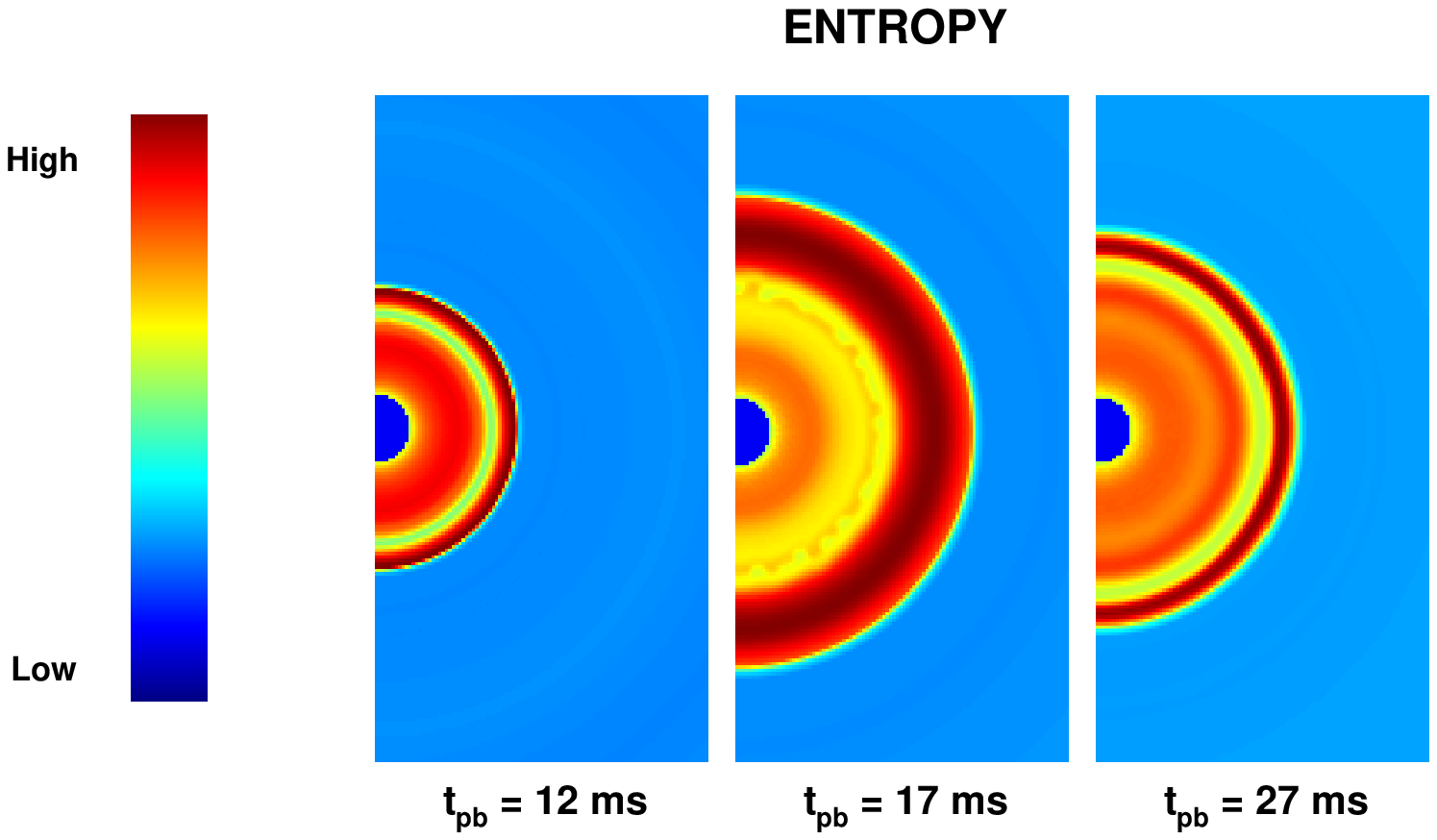,width=5.7in,angle=0}
\caption[Two-dimensional Entropy (Simulation G)]
{\protect\small \protect\baselineskip 10pt \label{pfig7a}
Two-dimensional plots showing the entropy evolution of the
25 ${\rm M}_{\odot}$  model in a ``low--test" simulation with
neutrino transport (simulation G).}
\end{figure}

\clearpage

% Figure page 15.

\addtocounter{numbercount}{1}

\begin{figure}[p]
\centering
\epsfig{file=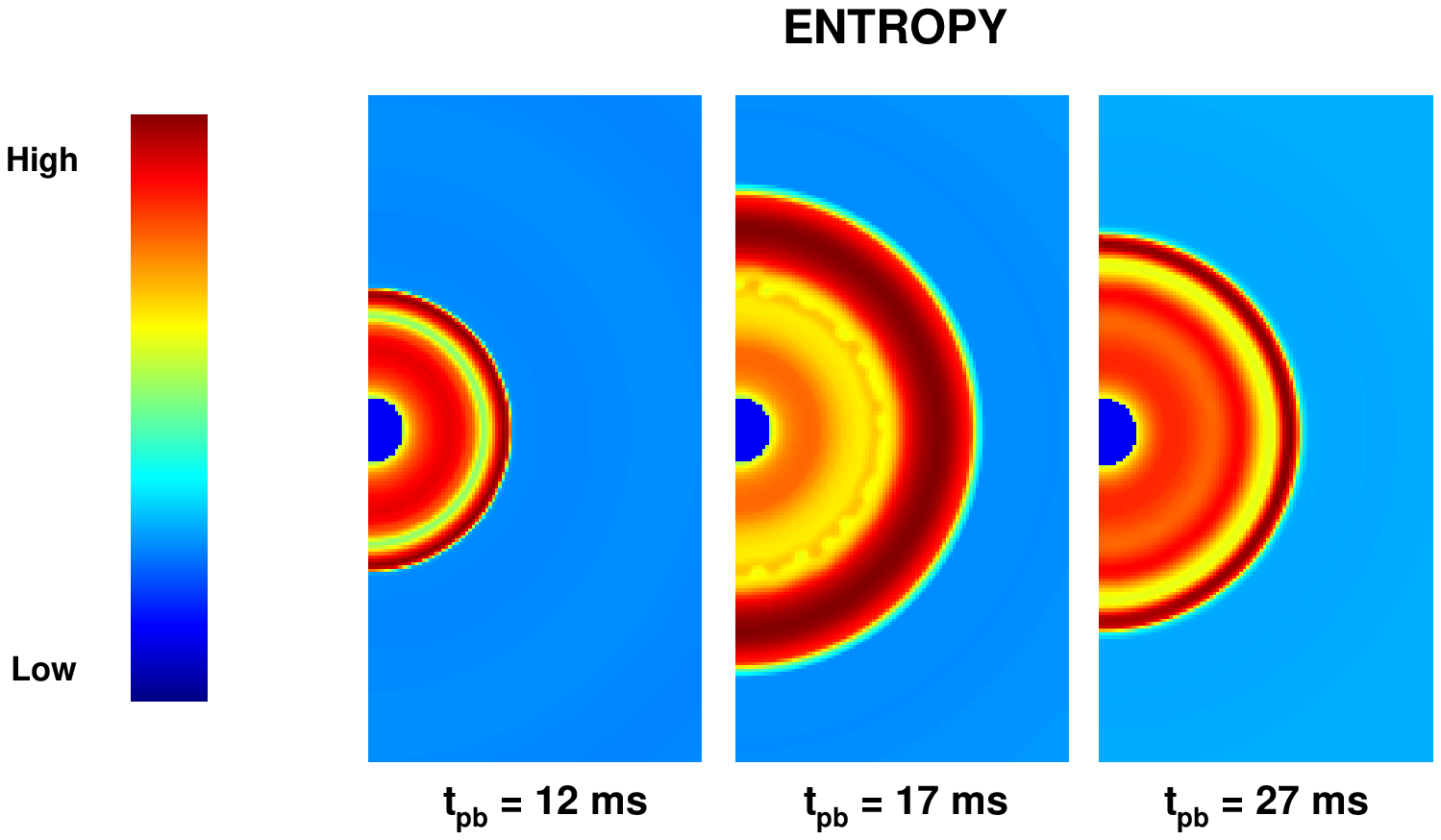,width=5.7in,angle=0}
\caption[Two-dimensional entropy (simulation H)]
{\protect\small \protect\baselineskip 10pt \label{pfig8a}
Two-dimensional plots showing the entropy evolution of the
25 ${\rm M}_{\odot}$  model in a ``low--test" simulation with
neutrino transport that includes ``finite-time" effects 
(simulation H).}
\end{figure}

\addtocounter{numbercount}{1}
\renewcommand{\thefigure}{\arabic{numbercount}}

\begin{figure}[p]
\centering
\epsfig{file=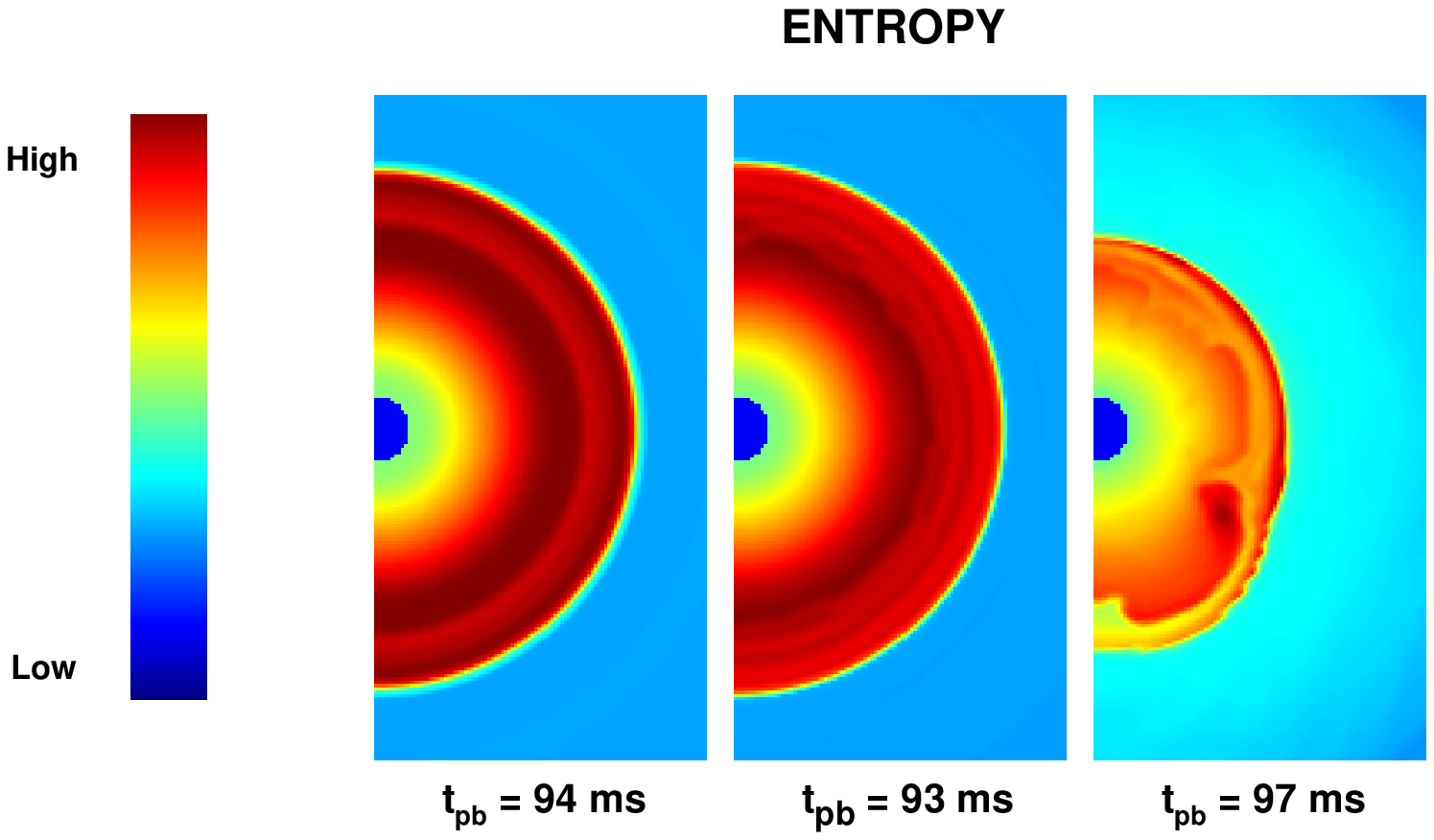,width=5.7in,angle=0}
\caption[Two-dimensional entropy (simulations D, E, and G)]
{\protect\small \protect\baselineskip 10pt \label{pfig9}
Two-dimensional entropy plots from simulations D, E, and 
G, at ``late" times.}
\end{figure}

\clearpage

% Figure page 16.

\addtocounter{numbercount}{1}
\renewcommand{\thefigure}{\arabic{numbercount}\alph{lettercount}}

\begin{figure}[p]
\centering
\epsfig{file=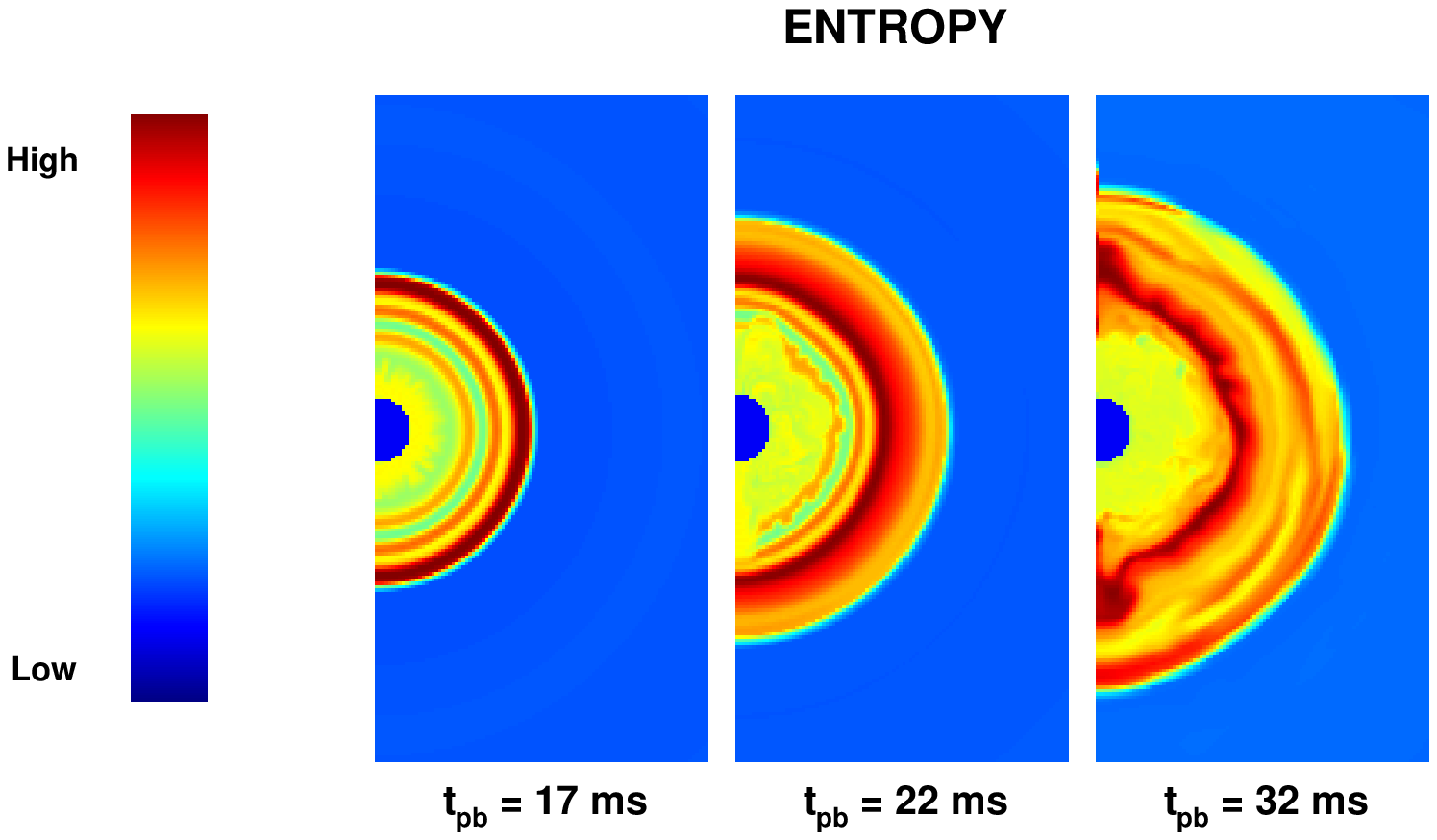,width=5.7in,angle=0}
\caption[Two-dimensional entropy (simulation K)]
{\protect\small \protect\baselineskip 10pt \label{pfig10a}
Two-dimensional plots showing the entropy evolution of the
15 ${\rm M}_{\odot}$  model in a simulation without neutrino 
transport (simulation K).}
\end{figure}

\addtocounter{numbercount}{1}

\begin{figure}[t]
\centering
\epsfig{file=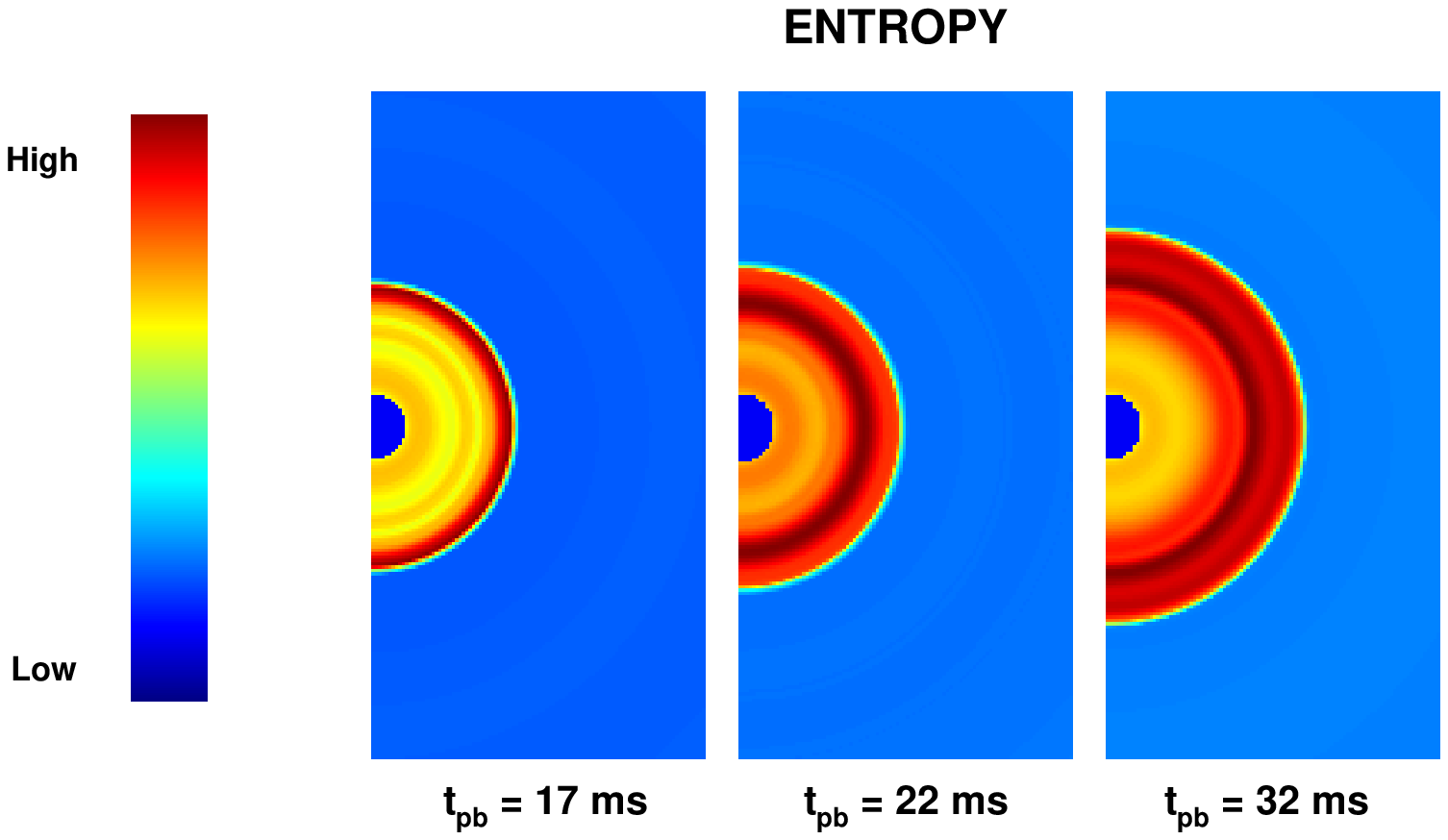,width=5.7in,angle=0}
\caption[Two-dimensional entropy (simulation L)]
{\protect\small \protect\baselineskip 10pt \label{pfig11a}
Two-dimensional plots showing the entropy evolution of the
15 ${\rm M}_{\odot}$  model in a simulation with neutrino
transport (simulation L).}
\end{figure}